\documentclass[fleqn,usenatbib]{mnras}

\usepackage{newtxtext,newtxmath}

\usepackage[T1]{fontenc}

\DeclareRobustCommand{\VAN}[3]{#2}
\let\VANthebibliography\thebibliography
\def\thebibliography{\DeclareRobustCommand{\VAN}[3]{##3}\VANthebibliography}

\usepackage{graphicx}	
\usepackage{amsmath}	
\usepackage{soul}

\title[Impacts of zonal winds on oscillations]
{Impacts of zonal winds on planetary oscillations and Saturn ring seismology}

\author[Dewberry, Mankovich, \& Fuller]{
Janosz W. Dewberry,$^{1,2}$\thanks{E-mail: dewberry@caltech.edu} Christopher R. Mankovich,$^3$ Jim Fuller$^{1}$
\\
$^{1}$ TAPIR, Walter Burke Institute for Theoretical Physics, Mailcode 350-17, California Institute of Technology, Pasadena, CA 91125, USA
\\
$^{2}$ Canadian Institute for Theoretical Astrophysics, 60 St. George Street, Toronto, ON M5S 3H8, Canada\\
$^{3}$ Division of Geological and Planetary Sciences, California Institute of Technology, Pasadena, CA 91125, USA
}

\date{Accepted 2022 July 07. Received 2022 July 07; in original form 2022 June 04}

\pubyear{2022}

\begin{document}
\label{firstpage}
\pagerange{\pageref{firstpage}--\pageref{lastpage}}
\maketitle

\begin{abstract}
The excitation of density and bending waves in Saturn's C ring by planetary oscillation modes presents a unique opportunity to learn about gas giant interiors and rotation. However, theoretical complications related to Saturn's rapid and differential rotation pose a barrier to the full utilization of ring wave detections. We calculate oscillation modes using a complete, non-perturbative treatment of differential rotation modelled after Saturn's zonal winds in self-consistently computed, polytropic equilibria. We find that previous, approximate treatments of the effects of differential rotation in Saturn overestimate shifts in the frequencies of fundamental modes (f-modes) thought to be responsible for the majority of the waves detected in the C ring, due to an omitted modification of the equilibrium shape and structure of the planet by differential rotation. The bias introduced by these frequency overestimates is small, but significant relative to the uncertainties afforded by Cassini data. We additionally consider the non-perturbative effects of Saturn-like differential rotation on the rotational mixing of f-modes and internal gravity modes (g-modes), which is relevant to detections of multiple density waves with very closely split pattern speeds. We find that higher order rotational effects can produce orders-of-magnitude enhancements in the surface gravitational perturbations of g-modes dominated by large spherical harmonic degrees $\ell$, regardless of frequency separation from the sectoral f-mode. Despite this enhancement, we find that the observed fine-splitting of density waves is unlikely to involve g-modes dominated by $\ell\gtrsim 10$. This restriction may aid in the inference of possible internal structures for Saturn.
\end{abstract}



\section{Introduction}
A subset of Saturn's seismic pulsations produce external, time-periodic gravitational perturbations with amplitudes sufficient to excite density and bending waves in Saturn's C ring \citep{Marley1991,Marley1993,Hedman2013,French2016,French2019,French2021,Hedman2019}. Identification of the specific internal oscillation modes responsible for external wave excitation provides insight into properties of Saturn's internal structure and rotation that are difficult to probe directly: \citet{Mankovich2019} used the identification of density waves characterized by higher azimuthal wavenumbers $m$ with excitation by Saturn's fundamental modes (f-modes) to place constraints on the planet's bulk rotation rate, which is otherwise difficult to measure because of close alignment between rotation and magnetic axes \citep{Cao2020}. \citet{Fuller2014b} and \citep{Mankovich2021} used observations of low-$m$ density waves, on the other hand, to make inferences regarding Saturn's deep interior, showing that an overabundance of $m=2$ and $m=3$ wave detections can be explained by internal gravity modes (g-modes) supported by a substantial region of stable stratification.

The fact that Saturn rotates rapidly poses a theoretical barrier to the accurate characterization of its oscillation modes. On top of a bulk rotation rate nearly $40\%$ of the dynamical frequency $\Omega_d=(GM_S/R_\text{eq}^3)^{1/2}$ (here $M_S$ and $R_\text{eq}$ are Saturn's mass and equatorial radius), the planet also exhibits latitude-dependant zonal winds \citep{Garcia-Melendo2011} that are thought to extend deep into the atmosphere \citep{Galanti2021}. Computing oscillation modes for the stably stratified interior models of \citet{Mankovich2021}, \citet{Dewberry2021} combined the first complete, ``non-perturbative'' treatment of the effects Saturn's rapid bulk rotation with a preliminary, approximate treatment of the slight differential rotation associated with these zonal winds. They found that Saturn's atmospheric zonal winds (i) produce measurable shifts in the frequencies of high-$m$ f-modes, and (ii) can marginally enhance rotational mixing of low-$m$ sectoral ($\ell\sim m$) f-modes with g-modes dominated by high spherical harmonic degrees $\ell$. The small asymmetry present in the observed winds is even capable of mixing g-modes and f-modes with different equatorial parities, producing equatorially asymmetric modes potentially capable of exciting both density and bending waves simultaneously.

Due to the latter enhanced rotational mixing, the approach of \citet{Dewberry2021} yielded sequences of modes with closely spaced frequencies and large surface gravitational perturbations (relative to mode energies). Such sequences provide a possible explanation for an observed $m=2$ doublet \citep{French2016} and an $m=3$ triplet \citep{Hedman2013} of density waves with frequencies separated by less than $1\%$, both of which are difficult to explain with fully perturbative treatments of Saturn's rapid bulk rotation \citep{Fuller2014a,Fuller2014b,Mankovich2019}. However, with their preliminary treatment of differential rotation, \citet{Dewberry2021} found that imbuing high-degree g-modes with gravitational perturbations sufficient for ring wave excitation required frequency separations that were in fact smaller than allowed by observational constraints.

In this paper, we reexamine both the f-mode frequency shifts and the f-mode/g-mode mixing considered by \citet{Dewberry2021} with an improved, fully non-perturbative treatment of differential rotation on cylinders in polytropic models. Focusing first on f-mode frequencies in $n=1$ polytropes with Saturn-like differential rotation, we recover qualitatively similar frequency shifts to \citet{Dewberry2021}. Quantitatively, though, a complete treatment of differential rotation leads to frequency shifts that can deviate from perturbative estimates by more than the observational uncertainty in ring wave pattern speed measurements. We attribute the deviations to changes in equilibrium structure that are omitted in the perturbative approach. 

Considering the low-$m$ modes of $n=1.6$ polytropes with a wide selection of parameterized regions of stable stratification, we also find that a fully non-perturbative treatment of differential rotation can lead to orders-of-magnitude enhancements in the surface gravitational perturbations of high-degree g-modes with frequencies close those of low-degree (sectoral) f-modes. This enhancement supports the ansatz that rotational mixing of sectoral f-modes with higher-$\ell$ g-modes may be responsible for the observed finely split $m=2$ and $m=3$ density waves. Despite this enhancement, we find that with realistic frequency separations from the f-modes, g-modes with eigenfunctions dominated by $\ell\gtrsim 10$ would encounter difficulty in exciting detectable density waves, unless preferentially excited to larger energies.

This paper is organized as follows. In \autoref{sec:num}, we describe the spectral methods used to compute differentially rotating, polytropic equilibria and their oscillation modes. \autoref{sec:res} then describes our results related to f-mode frequencies (\autoref{sec:fshift}) and rotational mode mixing (\autoref{sec:gmix}). Finally, we conclude in \autoref{sec:conc}.

\section{Numerical Methods}\label{sec:num}
\subsection{Rapidly rotating equilibria}\label{sec:eqm}
This subsection introduces our method for self-consistently computing oblate, differentially rotating equilibria. We are certainly not the first to calculate the structures of rapidly and differentially rotating polytropes \citep[see, e.g.,][]{Hachisu1986}, but for clarity we lay out our particular approach to this free-boundary problem. Those uninterested in the technical details of our model calculations may skip to \autoref{sec:rprof}. 

\subsubsection{Generalized Lane-Emden equation} 
Denoting equilibrium quantities by subscript $0'$s, we consider velocity fields with the form ${\bf u}_0={\bf \Omega}\times{\bf r}=R\Omega\hat{\boldsymbol{\phi}}$, where ${\bf \Omega}=\Omega(R)\hat{\bf z}$ (with $R=r\sin\theta$ the cylindrical radius). For such ``rotation on cylinders,'' the steady ($\partial_t=0$) Euler equation can be written as 
\begin{equation}\label{eq:hstat}
    \frac{\nabla P_0}{\rho_0}
    =-\nabla(\Phi_0+\Phi_\text{rot}),
\end{equation}
where $P_0$ is the equilibrium pressure, $\rho_0$ is the equilibrium density, $\Phi_0$ is the gravitational field, and 
\begin{equation}
    \Phi_\text{rot}
    =-\int_0^RR'\Omega^2(R')\text{d}R'
\end{equation}
is an effective centrifugal potential. 

\autoref{eq:hstat} must be supplemented by Poisson's equation $\nabla^2\Phi_0=4\pi G\rho_0,$ and an equation of state. For simplicity we consider polytropes characterized by $P_0\propto \rho_0^{1+1/n}$. Defining the pseudo-enthalpy $H =\int \text{d} P_0/\rho_0 =(1+n)P_0 /\rho_0\propto\rho_0^{1/n}$ then permits the direct integration of the Euler equation:
\begin{equation}\label{eq:Halg}
   \frac{H }{H_c}
   =\left(\frac{\rho_0}{\rho_c}\right)^{1/n}
   =1+\frac{1}{H_c}\left(
        \Phi_c 
        -\Phi_0
        -\Phi_\text{rot}
    \right),
\end{equation}
where subscript $c$'s denote central values. Noting that the boundary condition $H=0$ at the surface implies $H_c=\Phi_\text{pol}-\Phi_c$ \citep{Hachisu1986}, where $\Phi_\text{pol}$ gives the value of the gravitational potential at the pole, we define the usual Lane-Emden variable
\begin{equation}
    \Theta
    =\left(
        \frac{\Phi_\text{pol}-\Phi_0}
        {\Phi_\text{pol}-\Phi_c}
    \right)
    =\frac{1}{H_c}(H + \Phi_\text{rot}).
\end{equation}
Writing $\tilde{\Phi}_\text{rot}=\Phi_\text{rot}/H_c,\tilde{H}=H/H_c$ and immediately dropping tildes, we additionally scale lengths by the equatorial radius $R_\text{eq}.$ Substituting $\Theta$ into Poisson's equation, which  simplifies to Laplace's equation exterior to the surface $H=\Theta-\Phi_\text{rot}=0$, then leads to the non-dimensional relation
\begin{align}\label{eq:ThEqn}
    \nabla^2\Theta
    &=\begin{cases}
        -\lambda^2 \left(
            \Theta
            -{\Phi}_\text{rot}
        \right)^n & \Theta\geq{\Phi}_\text{rot} \\
        0 & \Theta<{\Phi}_\text{rot} 
    \end{cases}.
\end{align}
Here $\lambda^2=4\pi G\rho_cR_\text{eq}^2/H_c$ is an eigenvalue that results from fixing $R_\text{eq}=1$ \citep[see, e.g.,][]{Boyd2011}. \autoref{eq:ThEqn} therefore constitutes a nonlinear eigenvalue problem for the eigenvalue $\lambda$ and axisymmetric (but not spherically symmetric) solutions $\Theta(r,\theta)$. Rotation complicates the problem primarily because the surface $r_s=r_s(\theta)$ defined by $H=0$ (equivalently $\Theta<\Phi_\text{rot}$) is not known a priori.

\subsubsection{Newton iteration}
We adapt the Newton-Kantorovich iteration of \citet{Boyd2011} to two dimensions: at each $k$th iteration, we search for small corrections to both $\Theta$ and $\lambda$, writing $\Theta_{k+1}(r,\theta)=\Theta_k(r,\theta)+\delta_k(r,\theta)$ and $\lambda_{k+1}=\lambda_k+\epsilon_k.$ Substituting this iterative ansatz into \autoref{eq:ThEqn} and linearizing in $\delta$ and $\epsilon$ produces
\begin{align}\label{eq:iter2d}
    2\lambda_k(\Theta_k-{\Phi}_\text{rot})^n\epsilon_k
    &+\left[
        \nabla^2
        +n\lambda_k^2(\Theta_k-{\Phi}_\text{rot})^{n-1}
    \right]\delta_k
    \\&\notag
    \simeq
    -\nabla^2\Theta_k
    -\lambda_k^2(\Theta_k-{\Phi}_\text{rot})^n,
\end{align}
where $\Theta\geq\Phi_\text{rot}$, and $\nabla^2\delta_k\simeq-\nabla^2\Theta_k$ where $\Theta<\Phi_\text{rot}$. Each iteration thus poses a linear problem ${\bf J}_k\cdot{\bf x}_k={\bf r}_k$ for the $k$th correction, ${\bf x}_k=[\epsilon_k,\delta_k(r,\mu)]^T$, with a forcing function provided by the residual of the differential equation, ${\bf r}_k$. 

\subsubsection{Collocation}
We expand $\Theta$ (and each $\delta_k$) in a tensor basis of Chebyshev polynomials $T_\text{n}$ and (zonal) ortho-normalized spherical harmonics $Y_{\ell}^m$:
\begin{equation}\label{eq:ThExp}
    \Theta_k(r,\theta)
    =\sum_{\ell=0}^\infty\sum_{\text{n}=0}^\infty
    c_{\text{n}\ell}T_\text{n}[x(r)]Y_{2\ell}^{m=0}(\theta).
\end{equation}
This expansion leads to a spectral representation of differential operators (in this case the Laplacian). The pseudospectral approach of collocation \citep{Boyd2001} then involves minimizing the residual of the differential equation at advantageously chosen collocation nodes.

We split the radial domain between $r/R_\text{eq}\in[0,1]$ and $r/R_\text{eq}\in[1,\infty]$. Together with the assumed rotation on cylinders, the non-dimensionalization $R_\text{eq}=1$ ensures that $r_s$ falls inside the former domain; the outer vacuum is included in order to enforce a vanishing gravitational potential as $r\rightarrow\infty$. In the interior domain we map radius $r$ to a Gauss-Lobatto grid of collocation points $x\in[1,-1]$ via $r=(1 - x)/2$, while for the exterior we use the mapping $r=(1-x)/(1+x)+1.$ In the latitudinal direction, we use a Gauss-Legendre grid of points $\mu=\cos\theta\in[0,1)$. 

\subsubsection{Boundary conditions}
The appropriate radial boundary conditions, which we enforce in both the initial guess for $\Theta$ and all subsequent corrections $\delta$ via boundary bordering, are
\begin{align}
    \Theta|_{r=0,\forall\theta}&=1,\\
    \partial_r\Theta|_{r=0,\forall\theta}&=0,\\
    \Theta|_{r=R_\text{eq},\theta=\pi/2}&=0.
\end{align}
The last condition arises because of our choice to cast the equations as a nonlinear eigenvalue problem with a fixed radial scale ($R_\text{eq}$). Splitting the radial domain additionally introduces two interface conditions:
\begin{align}
    \Theta|_{r=R_\text{eq},\forall\theta}^+
    &=\Theta|_{r=R_\text{eq},\forall\theta}^-,\\
    \partial_r\Theta|_{r=R_\text{eq},\forall\theta}^+
    &=\partial_r\Theta|_{r=R_\text{eq},\forall\theta}^-,
\end{align}
where $-$ and $+$ refer to evaluations in the interior and exterior grids, respectively. Lastly, the physical boundary condition that the gravitational potential $\Phi_0\rightarrow0$ as $r\rightarrow\infty$ implies $\partial_r\Theta\rightarrow0$; Chebyshevs automatically satisfy this constraint under our adopted mapping between $x\in[1,-1]\rightarrow r\in[1,\infty]$.

\subsubsection{Comparison with previous work}
Our approach of solving a single nonlinear equation differs from ``self-consistent field'' (SCF) methods involving an iteration between the Euler and Poisson equations \citep[e.g.,][]{Ostriker1968,Hachisu1986,Jackson2005}. We also treat Poisson's equation in differential, rather than integral form. Integral form is usually favored \citep[e.g.,][]{Eriguchi1985,Hubbard2013} because it automatically ensures the application of the correct boundary condition as $r\rightarrow\infty$, but we find that iteratively computing two-dimensional quadratures at each grid point is more numerically expensive than adding an exterior computational domain. A third difference between our and many previous calculations is that we fix the rotation profile (in units of $\sqrt{H_c}/R_\text{eq}$) and allow the polar radius $R_\text{pol}$ to vary, rather than fixing the ratio $R_\text{pol}/R_\text{eq}$ and iterating to find a rotation profile \citep[e.g.,][]{Hachisu1986}. 

The generalization of \citet{Boyd2011} described in this section is most similar to the variational approach of \citet{Rieutord2016}, except that we do not find it necessary to evolve the grid while computing equilibrium structures. Despite the differences in our approach, our model computations agree with previously published results \citep[e.g.,][]{Passamonti2009,Passamonti2015} to the precisions reported. Appendix \ref{app:model} presents results from model calculations up to the mass-shedding limit for a range of polytropic indices.

\subsection{Rotation profiles}\label{sec:rprof}
Motivated by Saturn's observed surface-level winds, we consider angular velocities ${\bf \Omega}=\Omega(R)\hat{\bf z}$ with the form
\begin{equation}\label{eq:Omprof}
    \Omega(R)
    =\Omega_b
    +\frac{1}{2}A\Omega_b\text{sinc}[s(R-1)]
    \left\{
        1 + \text{tanh}
        \left[
            W(R - 1 + d)
        \right]
    \right\},
\end{equation}
where cylindrical $R=r\sin\theta$ is in units of $R_\text{eq}$, $s=7\pi, W=50,$ $d$ describes a decay depth (also in units of $R_\text{eq}$), and $A$ provides an amplitude relative to the bulk rotation rate $\Omega_b$ of the deep interior.

In altering $A$ and $d,$ we modify $\Omega_b$ relative to $\sqrt{H_c}/R_\text{eq}$ (the frequency unit of our model computations) in order to keep a fixed value of $\Omega(R=0)$ relative to the dynamical frequency $\Omega_d=(GM/R_\text{eq}^3)^{1/2}$ (the frequency unit of our oscillation computations). This choice stems from a desire to focus on the effects of differential, rather than bulk rotation on the oscillation modes, and from the fact that total mass and equatorial radius are measurable quantities for, e.g., Saturn. As an alternative, we have also computed modes for models in which we conserve the total angular momentum; for values of $A$ appropriate to Saturn, this approach yields similar results for f-mode frequencies.

\begin{figure*}
    \centering
    \includegraphics[width=\textwidth]{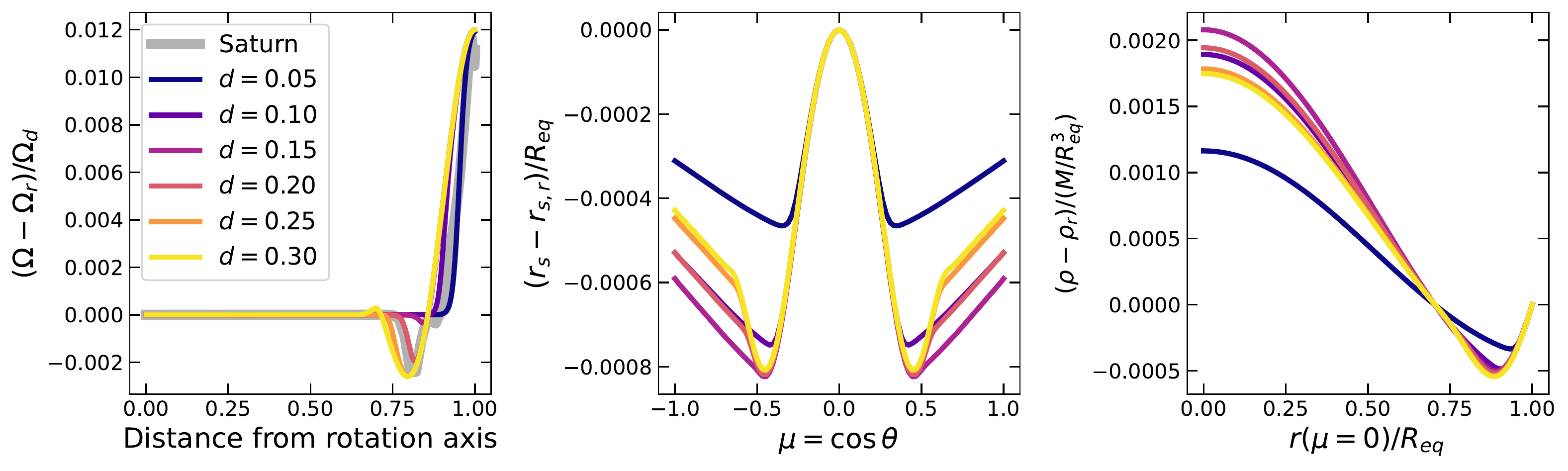}
    \caption{Left: difference between rotation profiles according to \autoref{eq:Omprof} (with $A=0.03,$ $\Omega_b/\Omega_d\simeq0.4$, and varying $d$), and a reference rigidly rotating model with constant rotation rate $\Omega_r/\Omega_{d}=\Omega_b/\Omega_d$. The faint gray line plots Saturn's differential rotation assuming the same decay with $d=0.2$. Middle: differences in surface radius as a function of $\mu=\cos\theta$ for the same models. Right: differences in equatorial density profiles. Note that the panels show deviations computed \emph{after} respectively re-scaling the differentially and rigidly rotating models to take on units with $G=M=R_\text{eq}=1$ \citep[this pre-normalization produces density and surface radius perturbations that differ in form from, e.g., those shown in figs. 7-8 of][]{Cao2017}.}
    \label{fig:Omprof}
\end{figure*}

\begin{figure*}
    \centering
    \includegraphics[width=\textwidth]{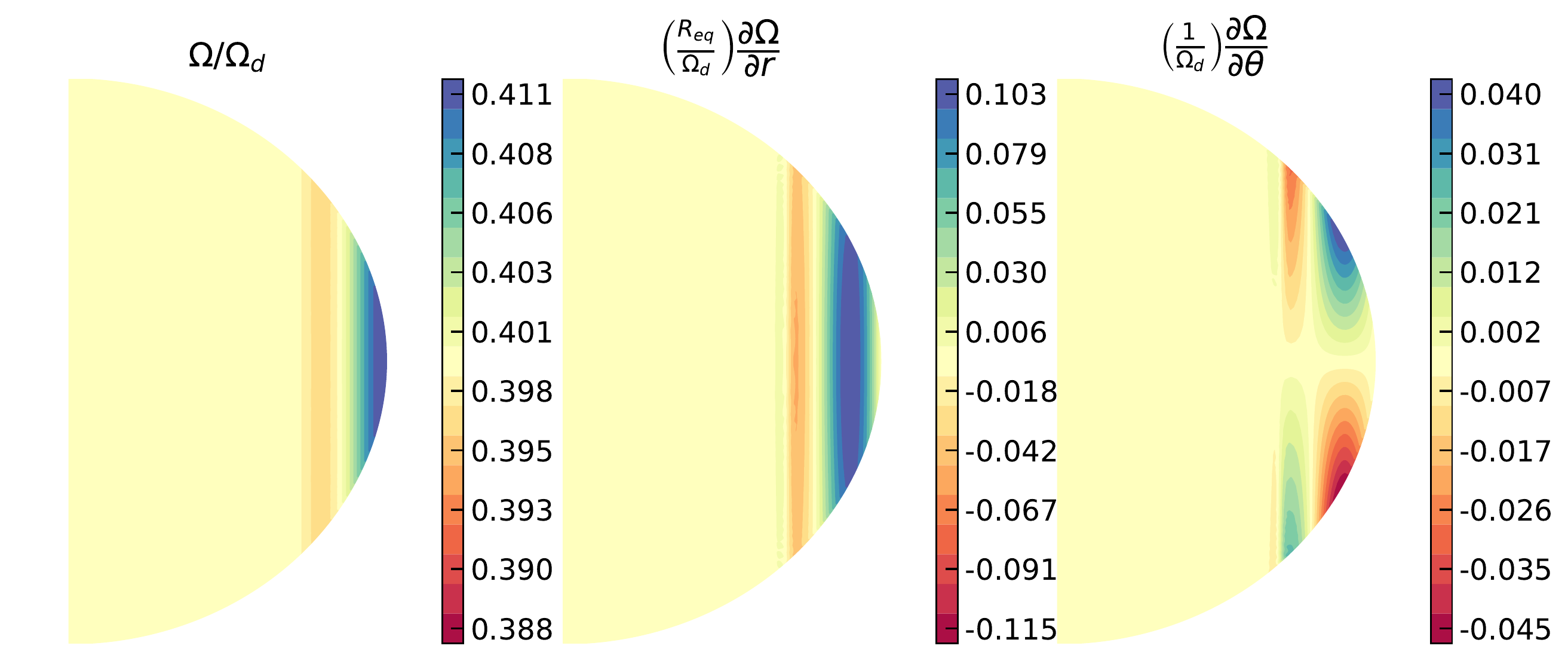}
    \caption{Cross-sections showing the meridional ($r,\theta$) structure of the differential rotation profile from \autoref{fig:Omprof} with $d=0.3$ (left) as well as partial derivatives with respect to spherical radius $r$ (middle) and colatitude $\theta$ (right).}
    \label{fig:Omslice}
\end{figure*}

In either case, changes in surface radius $r_s$ due to differential rotation necessarily imply a change in central density for the same total mass. This translates to a change in the density distribution (in units of $M/R_\text{eq}^3$) when computing modes. 
The panels in \autoref{fig:Omprof} show deviations in the equilibrium quantities of differentially rotating, $n=1$ polytropes from a reference rigidly rotating equilibrium with rotation rate $\Omega_r=\Omega(R=0)\simeq0.4\Omega_d$, for $A=0.03$ and different values of the depth parameter $d.$ The left panel shows differences in rotation as a function of $R$, with Saturn's rotation profile superimposed for an assumption of the same cylindrical decay with $d=0.2$. Meanwhile the middle panel plots differences in surface radius against $\mu=\cos\theta$, and the right panel shows deviations in equatorial density profiles (each normalized by $M/R_\text{eq}^3$ for the given model). The meridional cross-sections (slices along the rotation axis) in \autoref{fig:Omslice} then illustrate the ($r,\theta$)-structure of the $d=0.3$ rotation profile (left) and its derivatives (middle, right).

\subsection{Stratification}
We introduce stratification solely via modification of the first adiabatic exponent, $\Gamma_1$, which does not come into our calculations of equilibrium structure. Specifically, we assume the functional form
\begin{equation}\label{eq:Gm1}
    \Gamma_1(\zeta)=
    \begin{cases}
        \gamma,\zeta>\zeta_s\\
        \gamma 
        + A_s\{
            1 
            -\cos[
                \pi(\zeta-\zeta_s)
                /\zeta_s
            ]
        \},\zeta<\zeta_s
    \end{cases}
\end{equation}
where $\gamma=1+1/n$, and $\zeta$ is a quasi-radial coordinate defined for numerical convenience to match the surface of the oblate planet \citep[][see Appendix \ref{app:modes}]{Bonazzola1998}. The parameters $A_s$ and $\zeta_s$ control the amplitude and width of the profile for the Brunt-V\"ais\"al\"a (buoyancy) frequency, given by
\begin{equation}\label{eq:Brunt}
    N^2={\bf G}\cdot
    \left(
        \nabla\ln\rho_0 
        - \frac{1}{\Gamma_1}\nabla\ln P_0
    \right)
    =|{\bf G}|^2\frac{\rho_0}{P_0}
    \left(\frac{n}{n+1}-\frac{1}{\Gamma_1}\right),
\end{equation}
where ${\bf G}=\rho_0^{-1}\nabla P_0$ is the effective gravitational acceleration. The second equality in Equation \eqref{eq:Brunt} is specific to polytropes, and highlights a shortcoming of our simplified approach to modelling Saturn; although Saturn's convective envelope compares favorably with an $n=1$ polytrope, the choice of a given $n$ places the limitations on the possible amplitude of $N^2$ in the deep interior, namely $N^2\leq n|{\bf G}|^2\rho_0P_0^{-1}(n+1)^{-1}$ (assuming $\Gamma_1>0$). In order to produce models with profiles of $N^2$ with deep-interior magnitudes comparable to those considered by \citet{Mankovich2021} and \citet{Dewberry2021}, for our calculations focused on f-mode and g-mode mixing we choose $n=1.6$. Since f-modes are sensitive to the density profile in the envelope, this leads to f-mode frequencies that are offset from observations for Saturn. When it comes to evaluating f-mode and g-mode mixing, however, frequency separations are more important than absolute frequencies. We therefore expect our results for mode mixing to extend to more realistic models.

\begin{figure}
    \centering
    \includegraphics[width=\columnwidth]{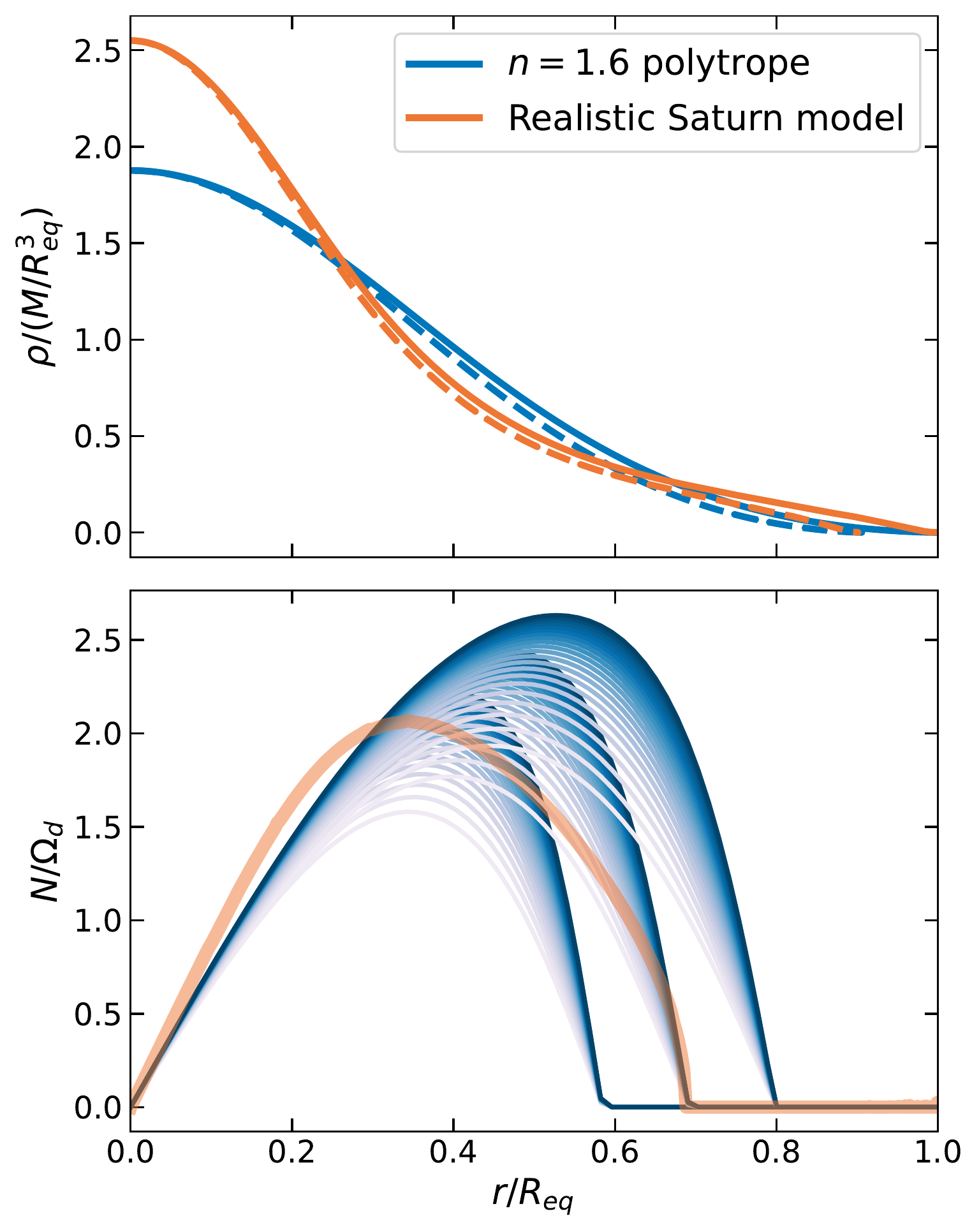}
    \caption{Top: equatorial (solid) and polar (dashed) profiles of density for a realistic (but rigidly rotating) model of Saturn (orange), and a differentially rotating, $n=1.6$ polytrope (blue) computed with $d=0.2$ and the same bulk internal rotation rate $\Omega_b/\Omega_d\simeq0.397$. Bottom: equatorial profiles of Brunt-V\"ais\"al\"a frequency produced by \autoref{eq:Gm1} with $\zeta_s=0.6,0.7,0.8$, and (from light to dark) increasing values of $A_s=2-10.$ The thick orange line shows the equatorial profile of the buoyancy frequency from the realistic, rigidly rotating Saturn model.}
    \label{fig:Brunt}
\end{figure}

\autoref{fig:Brunt} (top) compares the density profile of a differentially rotating, $n=1.6$ polytrope against that of a more realistic model of Saturn \citep[specifically, the fiducial model of ][]{Dewberry2021} with the same internal rotation rate, while \autoref{fig:Brunt} (bottom) shows a variety of equatorial profiles for the Brunt-V\"ais\"al\"a produced by setting $\zeta_s=0.6,0.7,0.8$ in \autoref{eq:Gm1}, and varying $A_s$ from $2$ (light) to $10$ (dark). The $n=1.6$ polytrope clearly does not involve as much central condensation as the more realistic model, but increasing the polytropic index further leads to significantly different f-modes. Future studies of the oscillations of differentially rotating planets should investigate piecewise polytropic, or more realistic equations of state.

\subsection{Oscillation mode calculations}
We compute oscillation modes using a non-perturbative treatment of the effects of both rapid and differential rotation. Introducing small-amplitude perturbations with the harmonic time-dependence $\exp[-\text{i}\sigma t]$ in the inertial frame, the linearizations of the equation of motion, continuity equation, energy equation, and Poisson's equation can be written as, respectively,
\begin{align}\label{eq:l1}
    D_t{\bf v}
    +{\bf v}\cdot\nabla{\bf u}_0
    &=-\rho_0^{-1}\nabla\delta P
    +\rho_0^{-1}{\bf G}\delta\rho
    -\nabla\delta\Phi,
\\
    D_t\delta\rho
    &=-\nabla\cdot(\rho_0{\bf v}),
\\
    D_t\delta P-c_A^2 D_t\delta\rho
    &={\bf v}\cdot(c_A^2\nabla\rho_0-\nabla P_0),
\\\label{eq:l4}
    0&=4\pi G\delta\rho-\nabla^2\delta\Phi.
\end{align}
Here ${\bf v},\delta P,\delta\rho,$ and $\delta\Phi$ are Eulerian perturbations to the velocity field, pressure, density, and gravitational field, and $c_A^2=\Gamma_1 P_0/\rho_0$ is the adiabatic sound speed. Lastly, we have written $D_t=\partial_t +{\bf u}_0\cdot\nabla=-\text{i}\sigma+{\bf u}_0\cdot\nabla$. 

The velocity field ${\bf u}_0$ breaks the spherical symmetry of the oscillation equations, through both its direct appearance and its modification of the equilibrium state. The linear eigenvalue problem posed by Equations \eqref{eq:l1}-\eqref{eq:l4} consequently involves non-separable partial differential equations, which we solve using spectral methods outlined by several previous authors \citep{Reese2006,Reese2009,Reese2013,Xu2017,Dewberry2021}. This approach involves expansions of the form 
\begin{equation}\label{eq:exp}
    \delta\Phi(r,\theta,\phi,t)=\mathcal{R}e\left(
        \exp(-\text{i}\sigma t)
        \sum_{\ell=m}^\infty\Phi^\ell(\zeta) Y_\ell^m(\theta,\phi)
    \right),
\end{equation}
where $\zeta$ is again a quasi-radial coordinate defined to match the oblate surface of the rotating fluid body. Appendix \ref{app:modes} provides the detailed expansion of Equations \eqref{eq:l1}-\eqref{eq:l4} in the (non-orthogonal) coordinate system $(\zeta,\theta,\phi)$, along with additional numerical details. Appendix \ref{app:barmode} provides validation via calculations of growth rates for dynamically unstable f-modes in highly distorted polytropes with ``constant-j'' profiles for differential rotation. 

In computing modes, we enforce the condition
\begin{equation}
    \langle
        \boldsymbol{\xi}_i,\boldsymbol{\xi}_i
    \rangle
    =\int_V\rho_0 
    \boldsymbol{\xi}_i^*
    \cdot\boldsymbol{\xi}_i
    \text{d}V=1,
\end{equation}
where $\boldsymbol{\xi}_i$ is the Lagrangian displacement of a mode with index $i.$ Although oscillation modes are not orthogonal under this inner product \citep[due to rotation; e.g., ][]{Schenk2002}, it still provides a convenient normalization.

\section{Results}\label{sec:res}
In this section we present the results of our mode calculations, focusing first on differential rotation's non-perturbative impact on f-mode frequencies, and second on its enhancement of mode mixing.


\subsection{Fundamental mode frequency shifts}\label{sec:fshift}
\begin{figure*}
    \centering
    \includegraphics[width=\columnwidth]{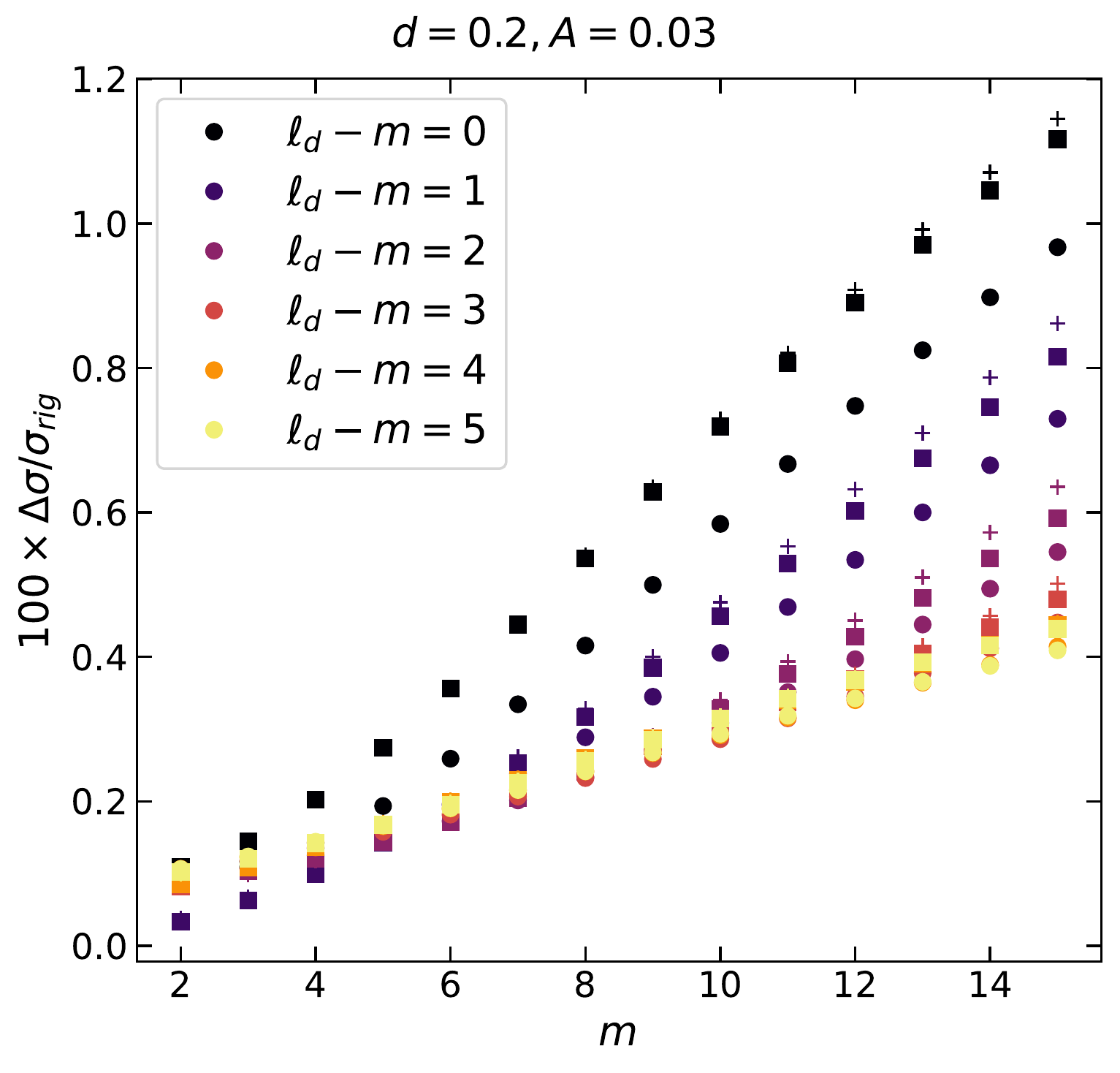}
    \includegraphics[width=\columnwidth]{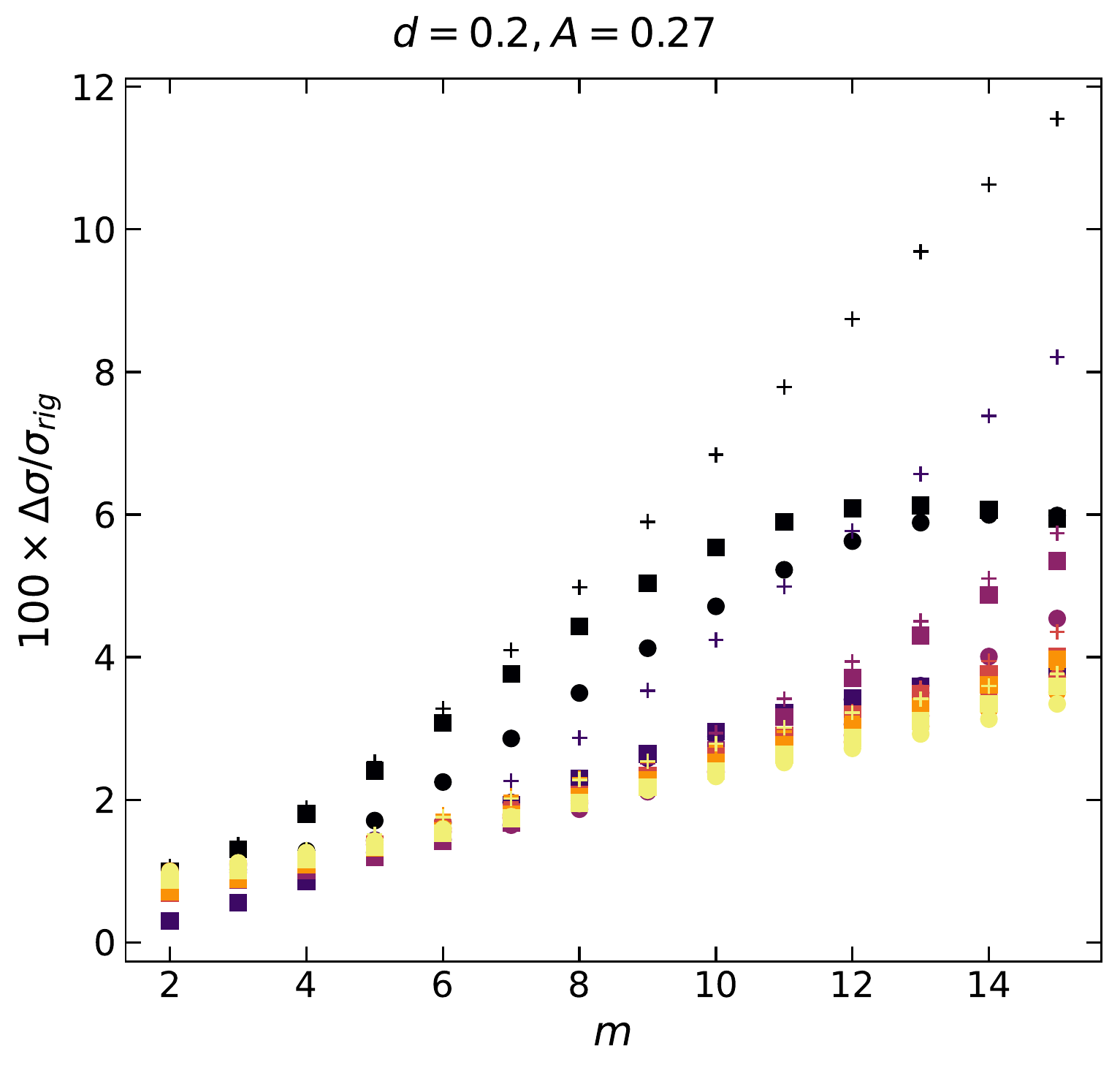}
    \caption{Left panel: Per cent changes in inertial-frame f-mode frequencies due to differential rotation described by by \autoref{eq:Omprof} with $\Omega_b/\Omega_d\simeq0.4,A=0.03,d=0.2$. Frequency shifts are plotted as a function of azimuthal wavenumber (x-axis), and dominant $\ell-m$ (colorscale). The filled circles show frequency shifts computed non-perturbatively, the plus signs show shifts computed with the perturbative treatment of differential rotation used by \citet{Dewberry2021}, and the squares show partially non-perturbative computations that (inconsistently) pair the differential rotation profile with the density/gravity/shape of the rigidly rotating model. The agreement between the squares and the plus signs indicates that the perturbative approach accurately captures the non-perturbative impact of the Coriolis force in this parameter regime; the disagreement with fully self-consistent calculations (circles) comes from the additional centrifugal flattening of the equilibrium by the zonal winds. Right panel: same as the left, but for modes computed from models with $d=0.2$ and a much larger $A=0.27$. For this case of much stronger differential rotation, the perturbative approach greatly exaggerates frequency shifts.}
    \label{fig:ddvar_1}
\end{figure*}

The panels in \autoref{fig:ddvar_1} plot per cent changes in (inertial-frame) f-mode frequencies due to differential rotation according to \autoref{eq:Omprof} (with  $\Omega_b/\Omega_d\simeq 0.4$), as a function of azimuthal wavenumber $m,$ for an $n=1$ polytrope. The left-hand panel shows results for Saturn-like depth and amplitude parameters $d=0.2,A=0.03$, while the right shows results for $d=0.2,A=0.27$. From dark to light, the point colors indicate increasing values of $\ell_d-m$, where $\ell_d$ is the degree of the dominant spherical harmonic in the oscillations' eigenfunctions. For example, $\ell_d-m=0$ corresponds to the ``sectoral'' f-modes with no zero-crossings in the polar direction, while modes with $\ell_d-m>0$ are the ``tesseral'' oscillations. 

The circles indicate frequency shifts computed with the full, non-perturbative treatment of differential rotation described in this paper. For comparison, the plus signs indicate frequency shifts calculated following the perturbative treatment of Saturn's zonal winds described in \citet{Dewberry2021} (see their eq. 24). The perturbative and non-perturbative calculations of frequency shifts look qualitatively similar in \autoref{fig:ddvar_1} (left), both to one another and to calculations employing more realistic, non-polytropic models of Saturn \citep[cf. fig. 7 of ][]{Dewberry2021}. Our calculations of frequency shifts also agree qualitatively with previous perturbative treatments of rotation in Saturn; \citet{Vorontsov1981} similarly observed an enhancement in the frequencies of prograde f-modes by differential (as opposed to purely rigid) rotation, finding as we do that this enhancement is most pronounced for the sectoral ($\ell_d-m=0$) oscillations (cf., the prograde modes in their fig. 2). 

However, the shifts computed with perturbative and non-perturbative treatments of rotation disagree quantitatively, most obviously for the sectoral modes. The disagreement is small, but surprisingly large relative to the small amplitude of the differential rotation ($\Omega-\Omega_b\simeq0.01\Omega_d$ at the surface). Importantly, the level of disagreement for the sectoral modes is much greater than the precision afforded by density and bending wave identifications in Saturn's rings \citep[$\lesssim 0.01$ per cent, corresponding to a 0.1 deg/d precision on a 1000 deg/d pattern speed;][]{French2021}.

The squares in \autoref{fig:ddvar_1} point toward the source of this discrepancy. For these calculations, we have solved Equations \eqref{eq:l1}-\eqref{eq:l4} in full with a given differential rotation profile (and its derivatives), while inconsistently assuming the equilibrium density, effective gravity, and surface radius of the reference rigidly rotating model with $\Omega_r=\Omega_b$ (i.e., we exclude \emph{additional} centrifugal flattening by differential rotation). This approximation is analogous to treatments of rigid rotation that completely include the Coriolis force, while excluding centrifugal distortion \citep[e.g.,][]{Takata2013}. The closer agreement of this partially non-perturbative approach with the perturbative shifts of \citet{Dewberry2021} (plus signs) therefore indicates that the deviation of both from the fully non-perturbative calculations originates in differential rotation's additional flattening of the background equilibrium. 

\begin{figure*}
    \centering
    \includegraphics[width=\columnwidth]{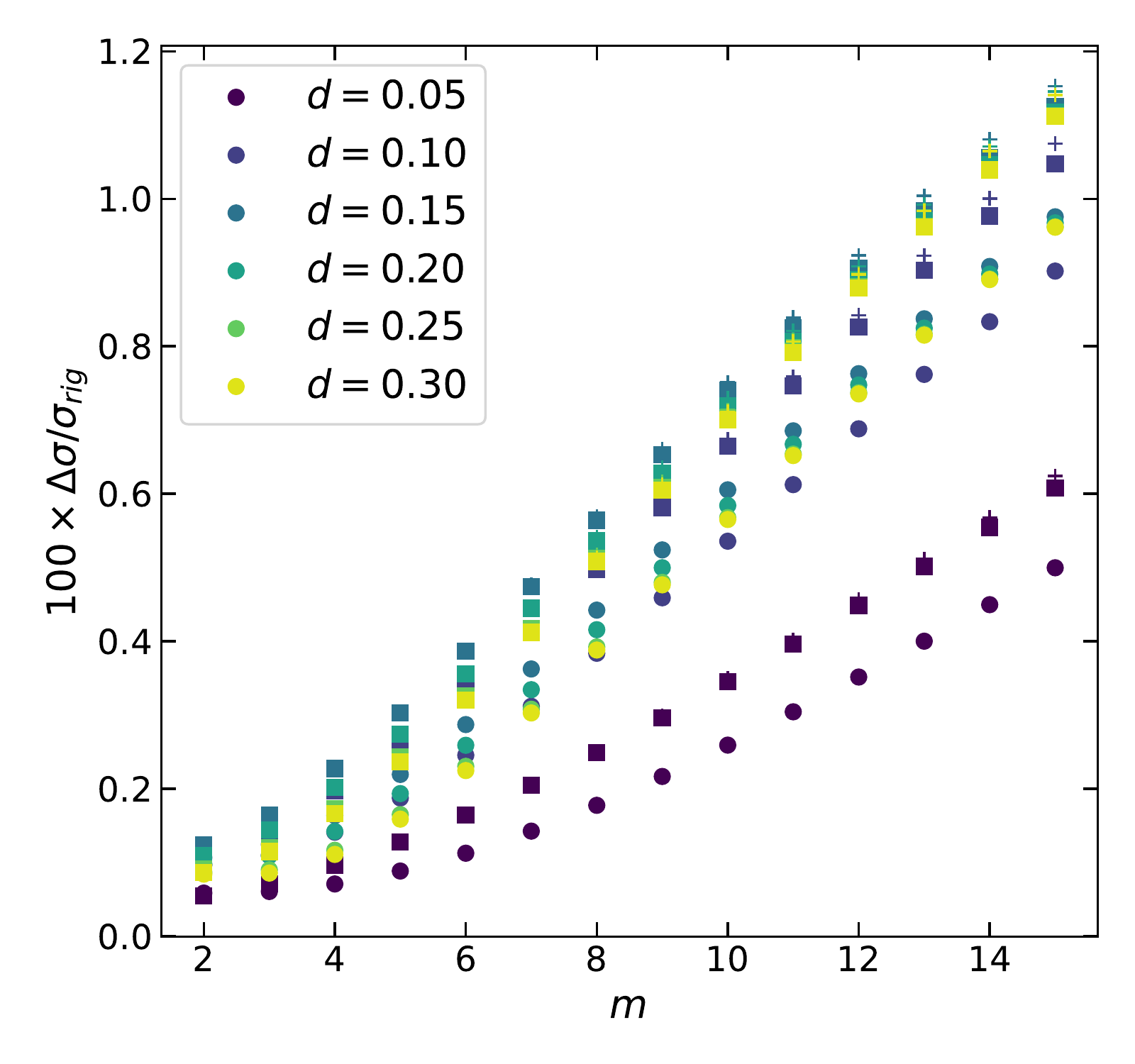}
    \includegraphics[width=\columnwidth]{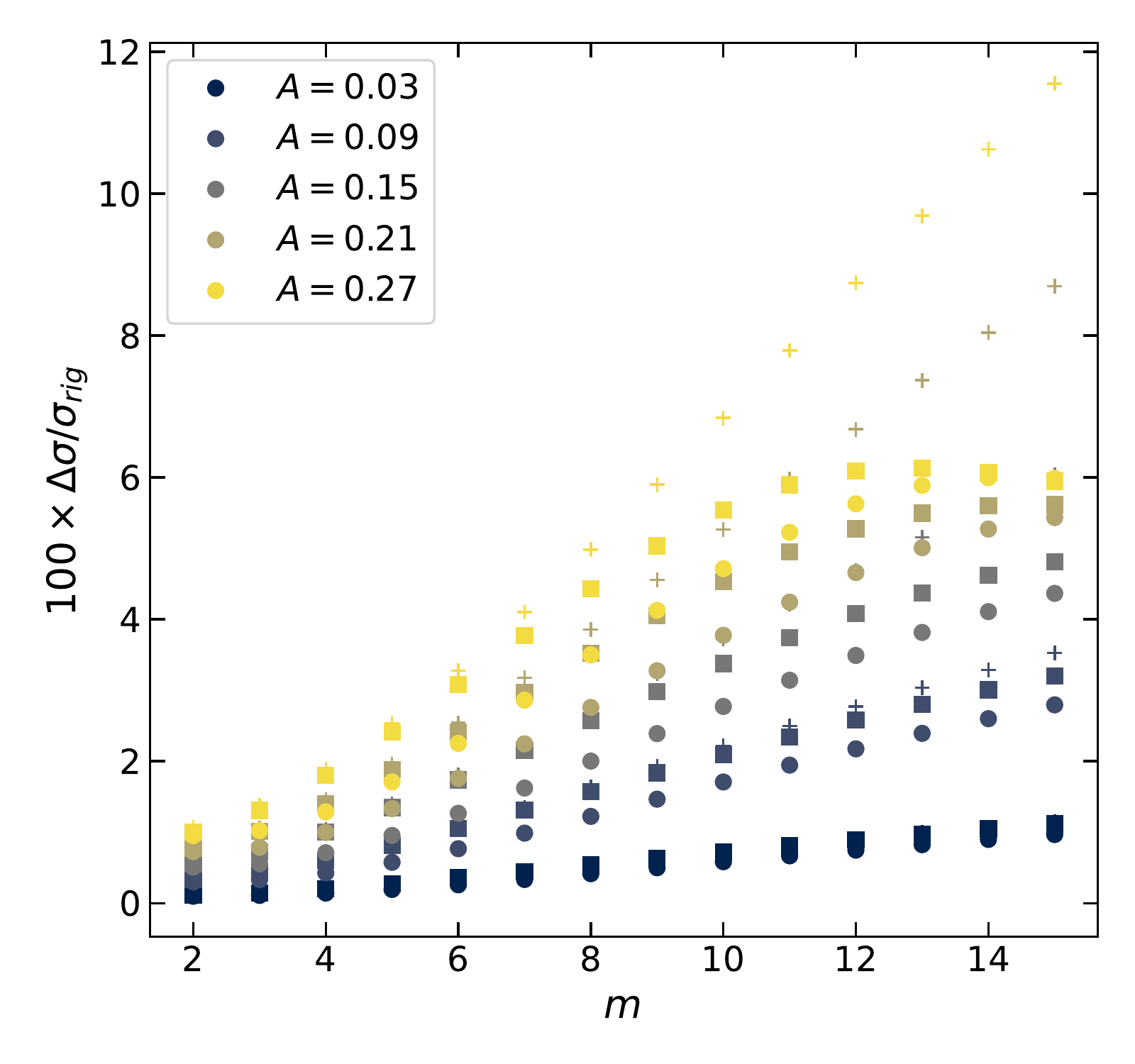}
    \caption{Frequency shifts for the \emph{sectoral} ($\ell_d-m=0$) f-modes, computed with varying $d,$ fixed $A=0.03$ (left), and varying $A$, fixed $d=0.2$ (right). Symbols have the same meaning as in \autoref{fig:ddvar_1}.}
    \label{fig:sectoralshifts}
\end{figure*}

\autoref{fig:ddvar_1} (right) plots the same f-mode frequency shifts computed for a much larger $A=0.27.$ This parameter value produces an equatorial jet with a surface amplitude $27\%$ larger than the rotation rate of the deep interior. For differential rotation this significant, the perturbative approach fails completely in capturing the properties of the sectoral f-modes, diverging significantly from both the partially and fully non-perturbative treatments.

\autoref{fig:sectoralshifts} focuses on the sectoral ($\ell_d-m=0$) f-modes. Both panels in the figure plot the same frequency shifts as \autoref{fig:ddvar_1}, as a function of increasing depth parameter $d$ (left) and amplitude parameter $A$ (right). Comparing with the rotation profiles in \autoref{fig:Omprof} (left), \autoref{fig:sectoralshifts} (left) indicates that the sectoral modes are most sensitive to the presence (or not) of a sub-corotating layer. Otherwise, the deviation of the perturbative and partially non-perturbative from the fully non-perturbative calculations appears relatively insensitive to the depth of wind decay, amounting to a maximum difference of $\simeq0.125\%$ ($0.175\%$) without (with) a sub-corotating layer for the $m=15$ sectoral f-mode. Meanwhile, the righthand panel illustrates the increasing inaccuracy of perturbative treatments for stronger and stronger differential rotation. 

\subsection{Mode mixing by differential rotation}\label{sec:gmix}
We now pivot to describe results related to mode mixing by differential rotation that may be relevant to observational identifications of $m=2$ and $m=3$ density waves with finely split frequencies in Saturn's rings \citep{Hedman2013,French2016}.

\subsubsection{Model ensemble}
\begin{figure*}
    \centering
    \includegraphics[width=\columnwidth]{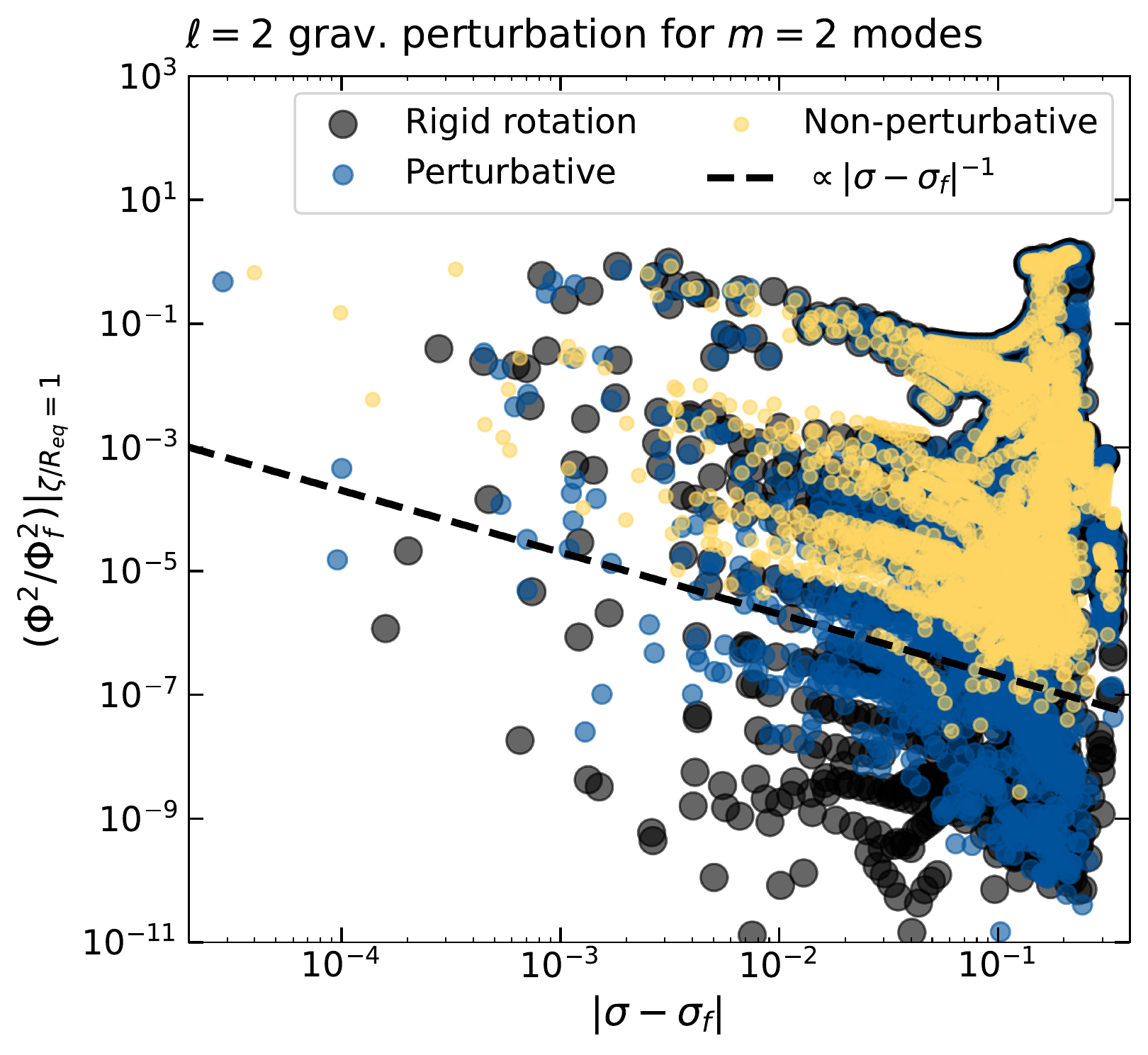}
    \includegraphics[width=\columnwidth]{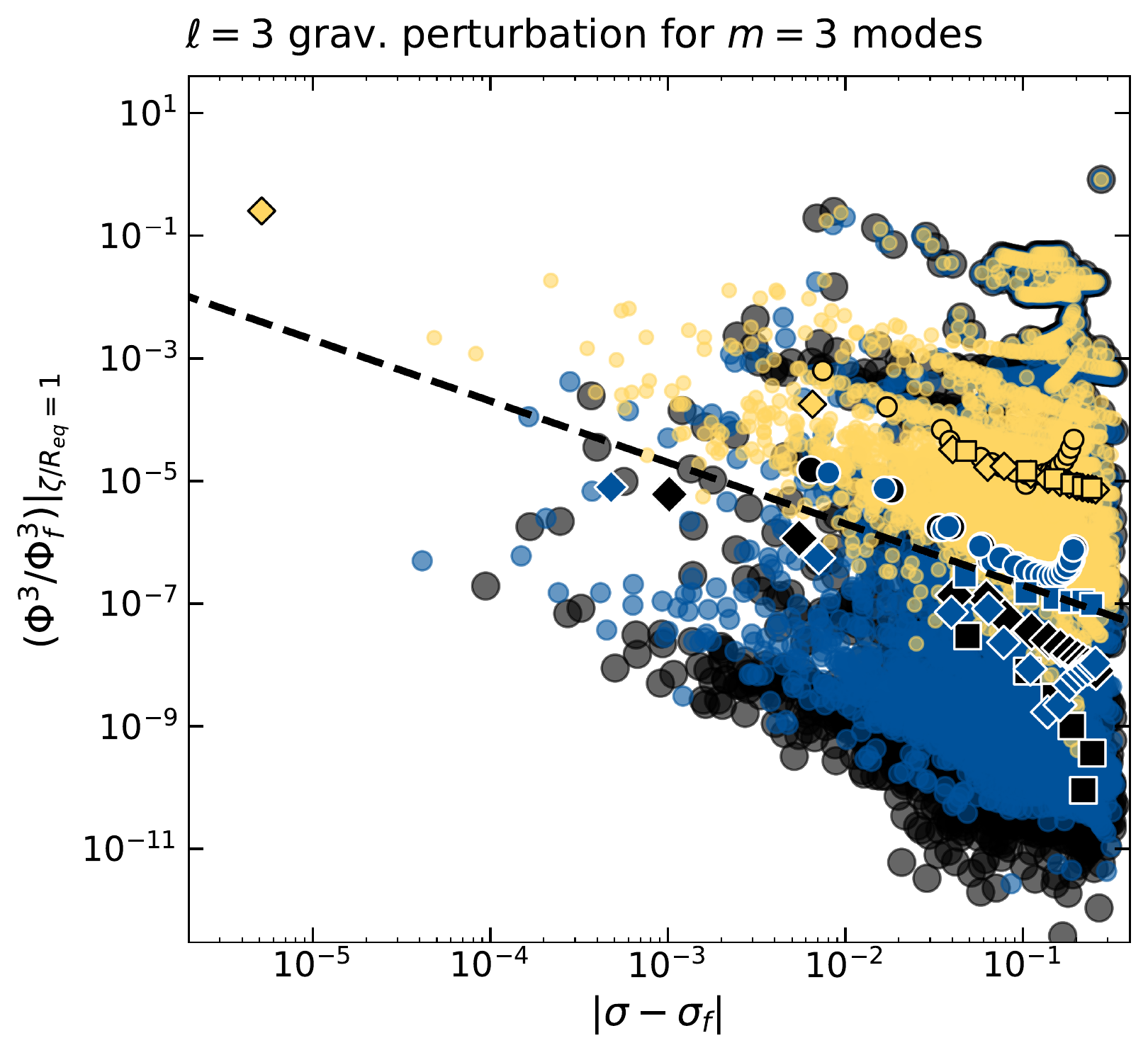}
    \caption{Surface value of the $\ell=m$ component of the gravitational perturbation of $m=2$ (left) and $m=3$ (right) modes computed for an ensemble of $n=1.6$ polytropes with Saturn's bulk rotation rate ($\Omega_b/\Omega_d\simeq0.397$), and regions of stable stratification determined by \autoref{eq:Gm1} with parameters ranging from $A_s=2$ to $10$, $\zeta_s/R_\text{eq}=0.6,0.7,0.8$. For each oscillation, the spectral amplitudes are normalized by, and plotted as a function of frequency separation (in $\Omega_d$) from the sectoral ($\ell_d=m$) f-mode with the same $m$. Black points show calculations including only rigid rotation, dark blue points show perturbative estimates of the effects of differential rotation (with $A=0.03,d=0.2$) computed with the approach described in \citet{Dewberry2021}, and gold points show fully non-perturbative calculations including the same zonal wind profile. The outlined points track the $m=3$ g-mode with dominant $\ell_d=17$ and radial order $\simeq3$ through models with different profiles of stratification (outlined circles, diamonds and squares respectively show calculations for models with $\zeta_s/R_\text{eq}=0.6,0.7,0.8$). As demonstrated by comparison with the black dashed lines, individual g-modes exhibit the inverse dependence $\Phi^m/\Phi^m_f\propto|\sigma-\sigma_f|^{-1}$ within frequency separations $|\sigma-\sigma_f|\lesssim0.1\Omega_d$ as the model properties vary.
    }
    \label{fig:mix}
\end{figure*}

\autoref{fig:mix} shows the results of a parameter survey of the ensemble of $n=1.6$ polytropic models shown in \autoref{fig:Brunt}. These models are characterized by (i) differential rotation according to \autoref{eq:Omprof} with fiducial parameters $A=0.03,d=0.2,\Omega_b/\Omega_d\simeq0.397$ \citep[Saturn's bulk rotation rate; ][]{Mankovich2019}, and (ii) internal regions of stable stratification determined by \autoref{eq:Gm1} with $\zeta_s/R_\text{eq}=0.6,0.7,0.8$ and $A_s=2-10$ (see \autoref{fig:Brunt}). For each model, we compute the $m=2$ and $m=3$ sectoral f-modes, and search for g-modes \citep[and ``rosette'' modes;][]{Takata2013} with similar frequencies. 

The points in both panels of \autoref{fig:mix} indicate surface values of the sectoral ($\ell=m$) spectral component $\Phi^m$ (see \autoref{eq:exp}) in the eigenfunctions for modes with $m=2$ (left) and $m=3$ (right). We plot each mode as a function of frequency separation from the respective sectoral f-mode, and normalize potential perturbations by those of the f-modes (denoted $\Phi^m_f$). The $\ell=m$ components of the gravitational perturbation are important because they decay least rapidly in the exterior vacuum, and therefore increase a given mode's ability to excite waves in external planetary rings. \citet{Dewberry2021} used a degenerate perturbative approach to estimate the enhancement of the low-degree components of the gravitational perturbations of g-modes via rotational mixing with f-modes by Saturn's zonal winds. The dark blue points in \autoref{fig:mix} show the results of applying the same perturbative method to modes computed from a reference rigidly rotating model (black points). On the other hand, the gold points show $\Phi^m$ values computed using a fully non-perturbative treatment of the same differential rotation profile.

\begin{figure*}
    \centering
    \includegraphics[width=\columnwidth]{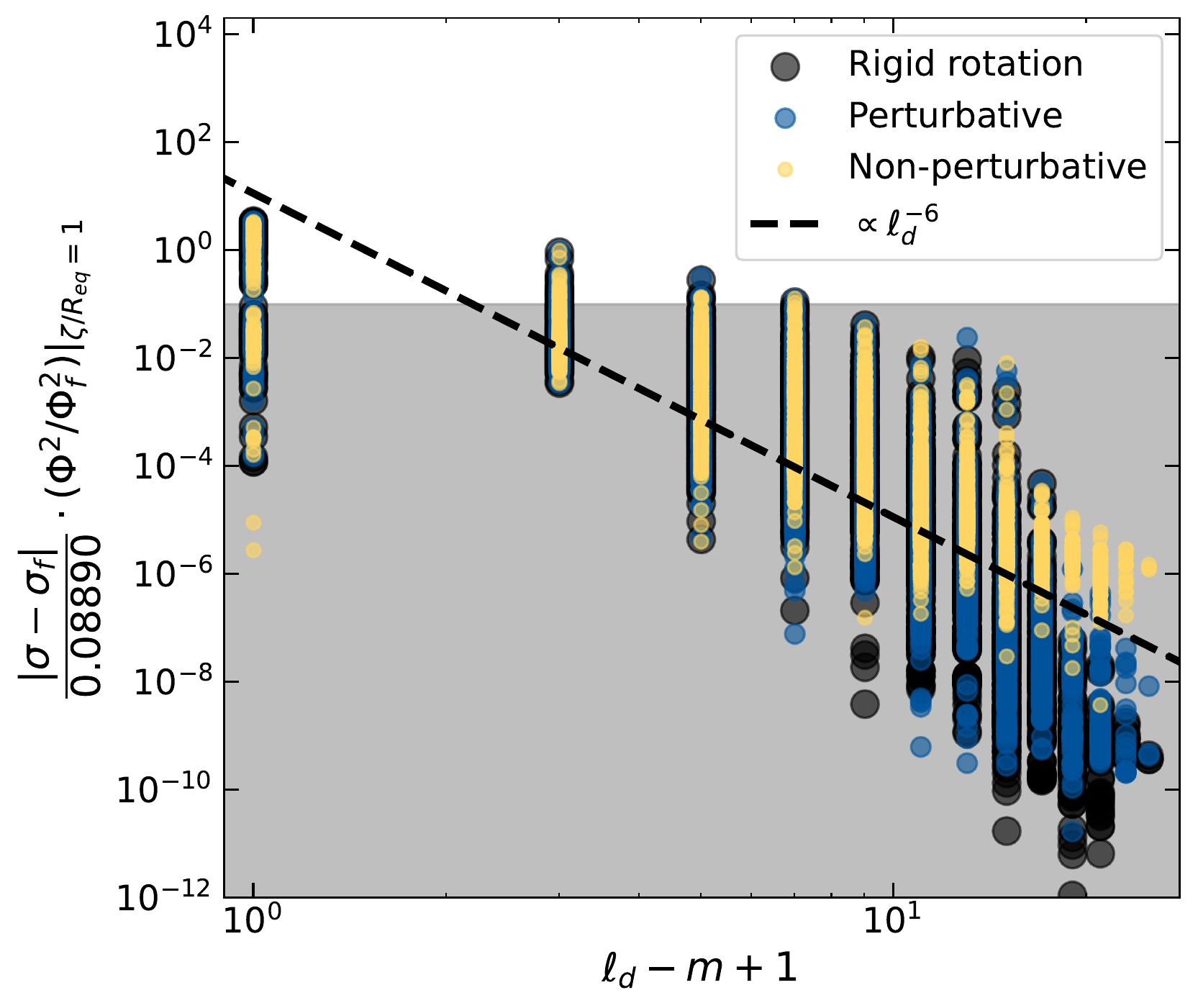}
    \includegraphics[width=\columnwidth]{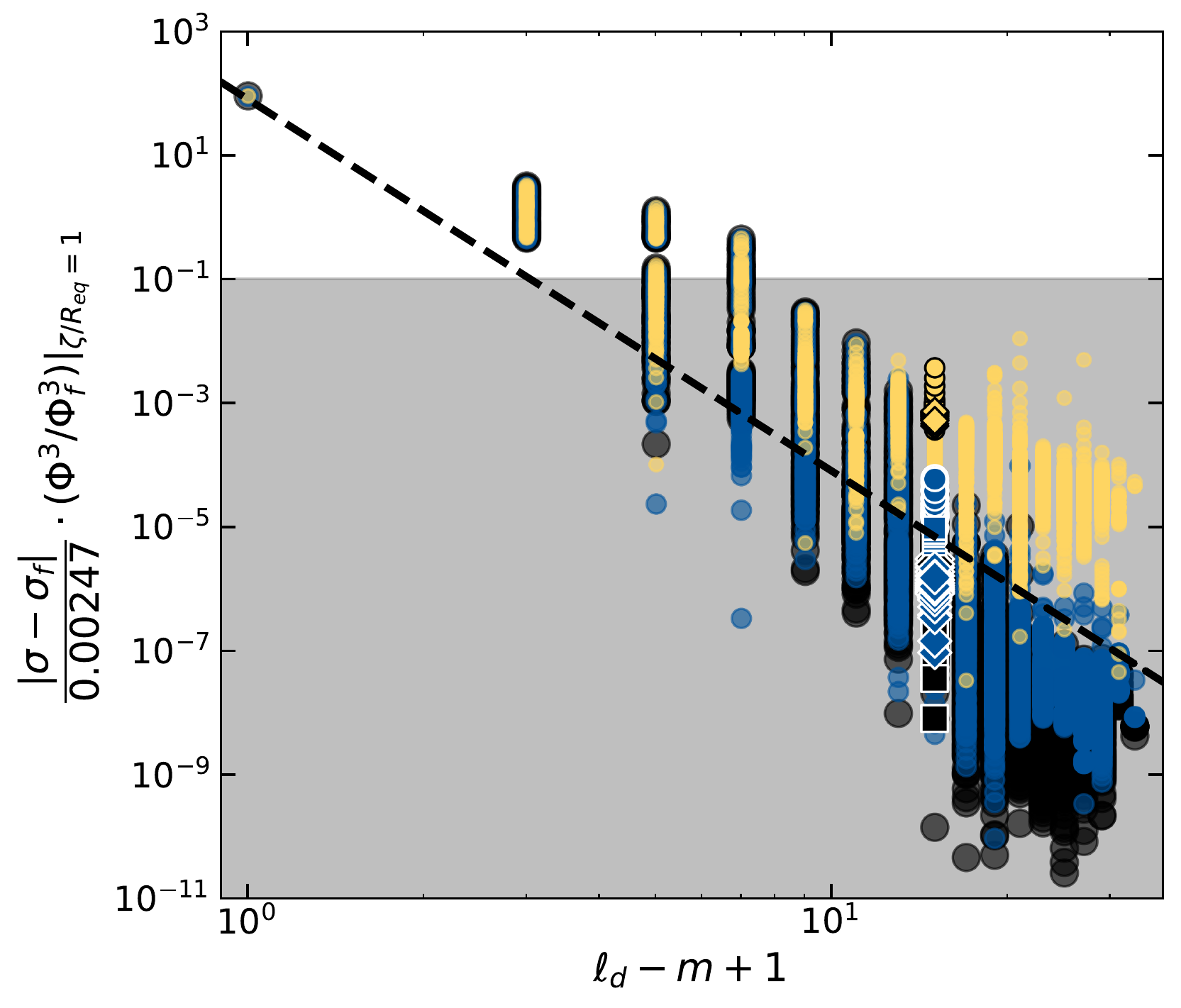}
    \caption{Plots showing surface $\Phi^m$ for the same $m=2$ (left) and $m=3$ (right) modes plotted in \autoref{fig:mix}, this time as a function of dominant degree $\ell_d$. The re-normalized y-values indicate the approximate potential perturbations each mode would have (relative to the f-mode) if its frequency separation from the f-mode were equal to the smallest frequency splitting observed for density waves in Saturn's rings with a given $m$. Values falling above $0.1$ indicate oscillations that could be reasonably expected to excite density waves consistent with the observations. Although higher-degree ($\ell_d\gtrsim m+10$) g-modes clearly gain gravitational enhancement from the fully non-perturbative treatment of differential rotation, they still do not fall within this detectable range, and therefore would not excite detectable density waves (unless driven to much larger energies than the f-modes).
    }
    \label{fig:mixl}
\end{figure*}

The outlined points show specific results for the $\ell_d=17,m=3$ g-mode that in the non-rotating limit would have radial order three (i.e., exactly $3$ zero-crossings in the radial direction). Outlined circles, diamonds and squares respectively show computations of this g-mode for models with $\zeta_s/R_\text{eq}=0.6,0.7$ and $0.8.$ The $\ell=m$ component of the gravitational perturbation for this oscillation, as well as all the others, shows an inverse dependence on frequency separation from the sectoral f-mode. Regardless of the treatment of rotation, tracking individual modes through the parameter space of stratification profiles reveals the inverse dependence 
\begin{equation}\label{eq:pwr}
    \left(\frac{\Phi^m}{\Phi^m_f}\right)\Bigg|_{\zeta/R_\text{eq}=1}
    \propto\frac{1}{|\sigma-\sigma_f|}
\end{equation}
within frequency separations $|\sigma-\sigma_f|\lesssim0.1\Omega_d.$ This inverse dependence agrees with expectations for mode mixing near avoided crossings \citep[see, e.g.,][]{Fuller2014a}. Although the power law manifests in all of our calculations, its \emph{prefactor} depends on the included effects of rotation: the smallest $\Phi^m$ computed non-perturbatively are orders of magnitude larger than the smallest amplitudes computed with purely rigid rotation, or the approximate treatment of differential rotation from \citet{Dewberry2021}. 

\subsubsection{Constraints on g-mode degrees}
In \autoref{fig:mixl}, we manipulate the data presented in \autoref{fig:mix} to highlight (i) which modes in particular are significantly affected by differential rotation, and (ii) how those effects relate to actual observations of $m=2$ and $m=3$ density waves in Saturn's rings. First of all, we utilize the dependence on frequency separation given in \autoref{eq:pwr} to compute scaled predictions of the $\Phi^m$ values that g-modes would possess \emph{if they had the closest frequency separations observed for ring waves.} Specifically, we identify the $m=2$ and $m=3$ density waves with pattern speeds $\Omega_p=\sigma/m\simeq1860.8 \, \text{deg/day}$ and $1735 \, \text{deg/day}$ (respectively) as excited by sectoral f-modes. This identification (which we note is not set in stone, particularly for $m=2$) then produces minimal frequency separations of $|\Omega_p-\Omega_{p,f}|\simeq91.6 \, \text{deg/day}\rightarrow|\sigma-\sigma_f|\simeq0.0889\Omega_d$ for the closest $m=2$ wave, and $|\Omega_p-\Omega_{p,f}|\simeq1.7 \, \text{deg/day}\rightarrow|\sigma-\sigma_f|\simeq0.0025\Omega_d$ for $m=3$. After normalizing the data in \autoref{fig:mix} by these values, and plotting as a function of dominant $\ell_d$, the panels in \autoref{fig:mixl} present estimates of the surface $\Phi^m$ that each g-mode mode would have \emph{if} the model were tuned so that the frequency separation from the f-mode matched the closest frequency splitting actually observed for each $m$. 

This normalization provides an imperfect mapping, as indicated by the spread in the outlined points, which again indicate calculations of the $\ell_d=17,m=3,n\sim3$ g-mode. In practice, repulsion near avoided crossings would also prevent some oscillations from coming so close in frequency to the f-mode, or from obtaining larger potential perturbations than the f-mode. Such repulsion should not impede the g-modes with higher $\ell_d,$ though, which exhibit the most interesting trends in \autoref{fig:mixl}. While all three treatments of rotation in Saturn give similar results for the low-$\ell_d$ g-modes, the fully non-perturbative treatment of differential rotation gives significantly larger surface $\Phi^m$ for those with $\ell_d\gtrsim m+10$ than the approximate approach of \citet{Dewberry2021}, which in turn gives marginally larger values than purely rigid rotation. \autoref{fig:mixl} thus indicates that ignoring differential rotation from Saturn's zonal winds entirely, or treating it perturbatively, leads to orders-of-magnitude underestimates of the gravitational enhancement of high-degree g-modes due to rotational mixing with f-modes. 

However, this enhanced mixing by differential rotation may not be enough to explain the observations. The white regions in \autoref{fig:mixl} indicate roughly the required surface gravitational perturbations for density wave excitation that would be detectable in Cassini data. The g-modes with surface $\Phi^m\lesssim0.1\Phi^m_f$ (i.e., those falling in the grey regions) would not excite detectable density waves in the rings, unless preferentially driven to larger energies than the f-modes by some mechanism. Consequently, \autoref{fig:mixl} suggests that, in the absence of preferential g-mode excitation, only g-modes with dominant $\ell_d\lesssim m + 10$ could be responsible for the observed density waves with the frequencies closest to the waves excited by the presumed f-modes. Since $\Phi^m/\Phi^m_f\propto|\sigma-\sigma_f|^{-1},$ stricter limits should apply to the observed density waves with larger frequency splittings.

\subsubsection{Representative example}
\begin{figure}
    \centering
    \includegraphics[width=\columnwidth]{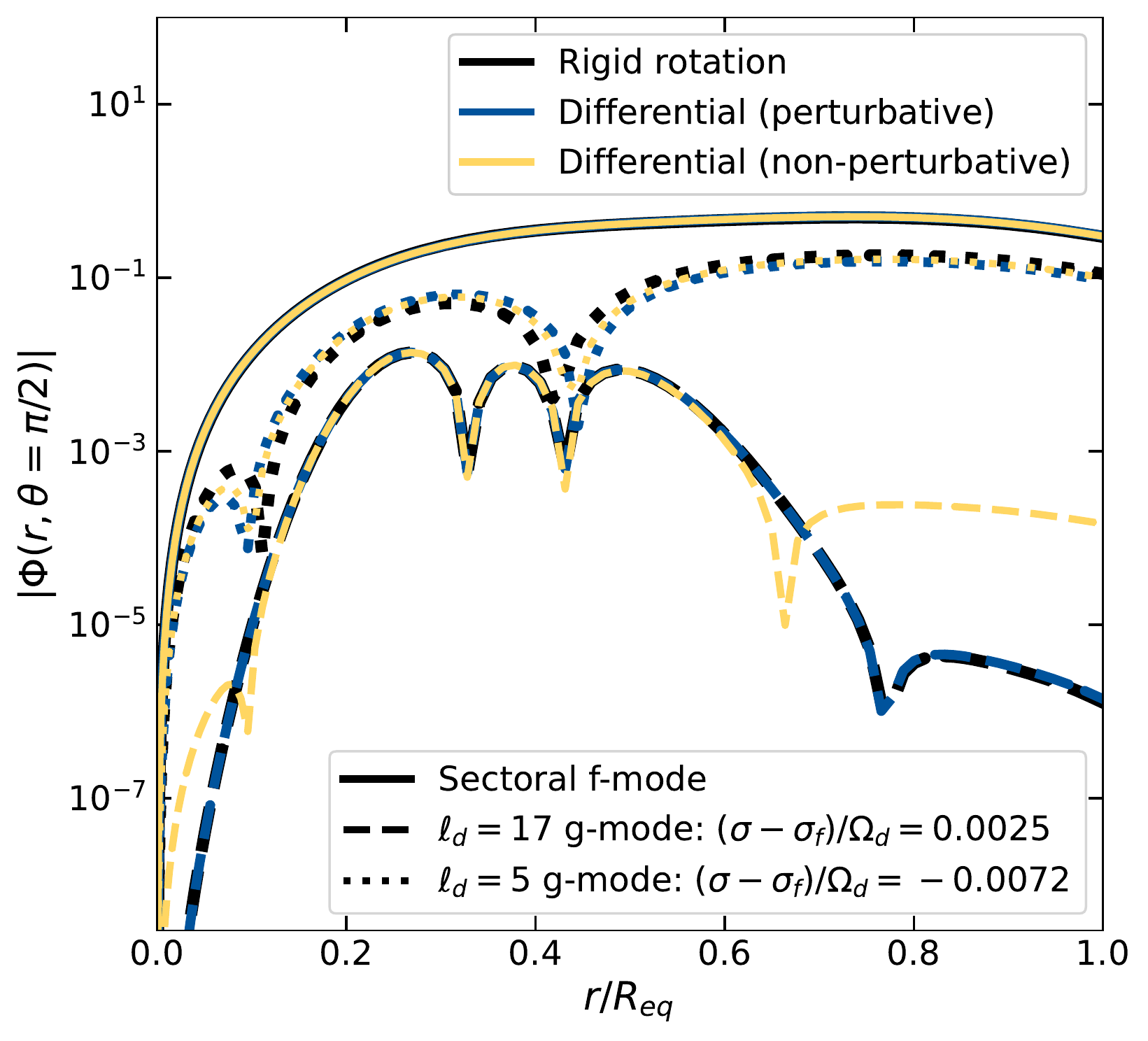}
    \caption{Equatorial radial profiles of gravitational perturbation for the $\ell_d=m=3$ f-mode (solid), an $\ell_d=5,m=3$ g-mode (dotted), and an $\ell_d=17,m=3$ g-mode (dashed) computed for an $n=1.6$ polytrope with differential rotation and stratification determined by \autoref{eq:Omprof} and \autoref{eq:Brunt} (with $d=0.2,A=0.03,\zeta_s/R_\text{eq}=0.7,A_s\simeq2.48$). As in \autoref{fig:mix} and \autoref{fig:mixl}, black, dark blue and gold lines respectively indicate calculations with purely rigid rotation, the approximate treatment of zonal winds from \citet{Dewberry2021}, and a fully non-perturbative treatment of differential rotation. Although the f-mode and $\ell_d=5$ g-mode eigenfunctions look similar for all three approaches, the non-perturbative treatment of differential rotation results in an enhancement of the surface gravitational perturbation of the $\ell_d=17$ g-mode by more than two orders of magnitude.}
    \label{fig:3mode}
\end{figure}

\autoref{fig:3mode} focuses on one illustrative example, plotting the (equatorial) radial profiles of gravitational perturbations computed for the sectoral $m=3$ f-mode (solid lines), and two g-modes with similar frequencies and $\ell_d=5$ (dotted) and $\ell_d=17$ (dashed). The gold lines show the results of fully non-perturbative calculations including our fiducial rotation profile with $d=0.2$ and $A=0.03,$ for a model with stratification determined by $\zeta_s/R_\text{eq}=0.7,A_s\simeq2.48$. Meanwhile, the black and dark blue lines respectively show calculations with purely rigid rotation, and the degenerate perturbative treatment of differential rotation from \citet{Dewberry2021}. For the calculations including differential rotation (both perturbatively and non-perturbatively), we have tuned the parameters so that the f-mode and $\ell_d=17$ g-mode are separated in frequency by $0.0025\Omega_d$ \citep[the frequency separation corresponding to the most closely spaced $m=3$ waves in Saturn's C ring;][]{Hedman2013}. 

The black and blue dashed lines in \autoref{fig:3mode} are barely distinguishable, which is not surprising given the modest difference between the black and blue points near the bottom of the panels in \autoref{fig:mix} and \autoref{fig:mixl}. In contrast, the non-perturbative treatment of differential rotation enhances the surface gravitational perturbation of the $\ell_d=17$ g-mode by more than two orders of magnitude. While it hardly affects the eigenfunction inside the stratified g-mode cavity, the non-perturbative mode mixing adds a low-$\ell$ component to the gravitational potential perturbation that prevents it from falling off steeply in the convective envelope. Meanwhile, all three treatments give comparable results for the f-mode and the $\ell_d=5$ g-mode.

The observed $m=3$ triplet in Saturn's C ring involves density waves with pattern speeds of $\Omega_p=\sigma/m\simeq1736.7,1735.0,1730.3\text{deg d}^{-1}$, and associated optical depth variations of $\delta\tau=0.07,0.21,0.15$ \citep[respectively;][]{Hedman2013}. For three modes with the same $m=3$ and nearly identical $\sigma,$ the $\sim1:3:2$ ratio of these optical depth perturbations translates roughly to a requirement of the same ratios between surface values of $\Phi^3$ under our normalization, additionally assuming energy equipartition between modes \citep[e.g.,][]{Fuller2014b}. The $\ell_d=17$ g-mode, f-mode, and $\ell_d=5$ g-mode plotted with gold lines in \autoref{fig:3mode} have surface $\Phi^3$ values in a ratio $\sim 0.001:3:1$, and so this particular mode interaction would struggle to explain the observations without preferential energy injection into higher-degree g-modes. 

Saturn's higher-latitude zonal winds (which we have excluded with our focus on barotropic rotation on cylinders) might further enhance rotational mixing. We do not expect these high-latitude winds to be particularly significant for the fine-splitting of $m=2$ and $m=3$ density waves, though, primarily because the eigenfunctions of the sectoral f-modes responsible for the g-modes' gravitational enhancement are confined relatively closely to the equator \citep[see, e.g., fig. 8 in ][]{Dewberry2021}. Consequently, Figures \ref{fig:mix}-\ref{fig:3mode} suggest that a robust explanation for the $m=3$ triplet in particular will likely require a more fortuitous near-degeneracy between the frequencies of the f-mode, and two g-modes with $\ell_d\lesssim13$. 

Given the large number of possible f-mode interactions with the dense spectra of g-modes produced by modern models of Saturn, this restriction to lower-degree oscillations should be helpful for observational inference. However, arranging the coincidence of frequencies for the $m=3$ f-mode and two lower-degree g-modes may require more complicated interior models than those considered in this paper, or by \citet{Dewberry2021}; despite spanning a wide region of parameter space, none of the models in the ensemble shown in \autoref{fig:Brunt} involve a near-frequency degeneracy between the $m=3$ f-mode and more than one relatively low-degree g-mode at a time. Multiple regions of stable stratification may provide greater flexibility for such three-mode interactions. Alternatively, the mixing of modes with different equatorial parities considered by \citet{Dewberry2021}, which we have excluded with our focus on barotropic rotation profiles in this work, might also allow for simpler reconciliation with $m=3$ observations.

\section{Conclusions}\label{sec:conc}
Saturn ring seismology presents one of our best observational windows into the interiors of giant planets. However, the planet's rapid and differential rotation stands in the way of full utilization of the ring wave data. In this paper, we have used a complete, non-perturbative treatment of differentially rotating polytropes to isolate the effects of Saturn-like differential rotation on (i) the frequencies of high-degree fundamental modes (f-modes), and (ii) the rotational mixing of low-degree f-modes with high-degree gravito-inertial modes (g-modes). 

First, we have shown that although approximate perturbative treatments of the effects of Saturn's differential rotation predict qualitatively similar f-mode frequency shifts to fully non-perturbative calculations, quantitatively the perturbative approach overestimates these shifts by up to $\simeq0.1-0.2\%$ of the inertial-frame mode frequency (see \autoref{fig:ddvar_1}, \autoref{fig:sectoralshifts}). In Saturn's case, this amounts to a roughly $10\%$ overestimate of the frequency shift due to differential rotation. We attribute the discrepancy to the perturbative omission of the modification of the equilibrium planet's shape and structure by differential rotation. Biases of $\simeq0.1-0.2\%$ are small, but much larger than the $\lesssim0.01\%$ uncertainty inherent to the wave detections in Cassini data \citep{French2021}.

We have additionally compared the mixing of f-mode and g-mode eigenfunctions due to the non-perturbative effects of differential rotation against estimates from the degenerate perturbative approach of \citet{Dewberry2021}. Although a perturbative treatment of Saturn's zonal winds is sufficient for relatively low-degree g-modes, we find that it drastically underestimates the surface gravitational perturbations of high-degree g-modes, at a given frequency separation from the f-mode  (\autoref{fig:mix}, \autoref{fig:mixl}, \autoref{fig:3mode}). 

This enhancement due to the non-perturbative effects of differential rotation is important to the search for a definitive explanation for observations of $m=2$ and $m=3$ density waves with frequencies split by less than one per cent \citep{Hedman2013,French2016}. Importantly, however, we find that even with an orders-of-magnitude enhancement due to differential rotation, g-modes dominated by spherical harmonic degrees $\ell\gtrsim m+10$ would still have surface gravitational perturbations too small to produce detectable density waves with the smallest frequency separations observed for $m=2$ and $m=3$, absent an excitation mechanism that preferentially excites high-degree g-modes to larger energies. This restriction to lower-degree g-modes may aid in limiting the subset of possible interior models for Saturn.

\section*{Acknowledgements}
We thank Mark S. Marley for reviewing this work, and for providing helpful comments that improved the quality of the paper. We are thankful for support from the Caltech Center for Comparative Planetary Evolution. JWD gratefully acknowledges support from the Sloan Foundation through grant FG-2018-10515, and from the Natural Sciences and Engineering Research Council of Canada (NSERC) [funding reference \#CITA 490888-16].

\section*{Data Availability}
The data underlying this work will be provided upon reasonable request to the corresponding author.



\bibliographystyle{mnras}
\bibliography{fmode_diffrot}

\begin{thebibliography}{}
\makeatletter
\relax
\def\mn@urlcharsother{\let\do\@makeother \do\$\do\&\do\#\do\^\do\_\do\%\do\~}
\def\mn@doi{\begingroup\mn@urlcharsother \@ifnextchar [ {\mn@doi@}
  {\mn@doi@[]}}
\def\mn@doi@[#1]#2{\def\@tempa{#1}\ifx\@tempa\@empty \href
  {http://dx.doi.org/#2} {doi:#2}\else \href {http://dx.doi.org/#2} {#1}\fi
  \endgroup}
\def\mn@eprint#1#2{\mn@eprint@#1:#2::\@nil}
\def\mn@eprint@arXiv#1{\href {http://arxiv.org/abs/#1} {{\tt arXiv:#1}}}
\def\mn@eprint@dblp#1{\href {http://dblp.uni-trier.de/rec/bibtex/#1.xml}
  {dblp:#1}}
\def\mn@eprint@#1:#2:#3:#4\@nil{\def\@tempa {#1}\def\@tempb {#2}\def\@tempc
  {#3}\ifx \@tempc \@empty \let \@tempc \@tempb \let \@tempb \@tempa \fi \ifx
  \@tempb \@empty \def\@tempb {arXiv}\fi \@ifundefined
  {mn@eprint@\@tempb}{\@tempb:\@tempc}{\expandafter \expandafter \csname
  mn@eprint@\@tempb\endcsname \expandafter{\@tempc}}}

\bibitem[\protect\citeauthoryear{{Bonazzola}, {Gourgoulhon}  \&
  {Marck}}{{Bonazzola} et~al.}{1998}]{Bonazzola1998}
{Bonazzola} S.,  {Gourgoulhon} E.,   {Marck} J.-A.,  1998, \prd, 58, 104020

\bibitem[\protect\citeauthoryear{{Boyd}}{{Boyd}}{2001}]{Boyd2001}
{Boyd} J.~P.,  2001, {Chebyshev and Fourier Spectral Methods}.
Dover Publications, Inc

\bibitem[\protect\citeauthoryear{{Boyd}}{{Boyd}}{2011}]{Boyd2011}
{Boyd} J.~P.,  2011, Numerical Mathematics: Theory, Methods and Applications,
  4, 142

\bibitem[\protect\citeauthoryear{{Cao} \& {Stevenson}}{{Cao} \&
  {Stevenson}}{2017}]{Cao2017}
{Cao} H.,  {Stevenson} D.~J.,  2017, \mn@doi [Journal of Geophysical Research
  (Planets)] {10.1002/2017JE005272}, \href
  {https://ui.adsabs.harvard.edu/abs/2017JGRE..122..686C} {122, 686}

\bibitem[\protect\citeauthoryear{{Cao}, {Dougherty}, {Hunt}, {Provan},
  {Cowley}, {Bunce}, {Kellock}  \& {Stevenson}}{{Cao} et~al.}{2020}]{Cao2020}
{Cao} H.,  {Dougherty} M.~K.,  {Hunt} G.~J.,  {Provan} G.,  {Cowley} S. W.~H.,
  {Bunce} E.~J.,  {Kellock} S.,   {Stevenson} D.~J.,  2020, \icarus, 344,
  113541

\bibitem[\protect\citeauthoryear{{Dewberry} \& {Lai}}{{Dewberry} \&
  {Lai}}{2022}]{Dewberry2022}
{Dewberry} J.~W.,  {Lai} D.,  2022, \mn@doi [\apj] {10.3847/1538-4357/ac3ede},
  \href {https://ui.adsabs.harvard.edu/abs/2022ApJ...925..124D} {925, 124}

\bibitem[\protect\citeauthoryear{{Dewberry}, {Mankovich}, {Fuller}, {Lai}  \&
  {Xu}}{{Dewberry} et~al.}{2021}]{Dewberry2021}
{Dewberry} J.~W.,  {Mankovich} C.~R.,  {Fuller} J.,  {Lai} D.,   {Xu} W.,
  2021, \mn@doi [PSJ] {10.3847/PSJ/ac0e2a}, \href
  {https://ui.adsabs.harvard.edu/abs/2021PSJ.....2..198D} {2, 198}

\bibitem[\protect\citeauthoryear{{Eriguchi} \& {Mueller}}{{Eriguchi} \&
  {Mueller}}{1985}]{Eriguchi1985}
{Eriguchi} Y.,  {Mueller} E.,  1985, \aap, \href
  {https://ui.adsabs.harvard.edu/abs/1985A&A...146..260E} {146, 260}

\bibitem[\protect\citeauthoryear{{French}, {Nicholson}, {Hedman}, {Hahn},
  {McGhee-French}, {Colwell}, {Marouf}  \& {Rappaport}}{{French}
  et~al.}{2016}]{French2016}
{French} R.~G.,  {Nicholson} P.~D.,  {Hedman} M.~M.,  {Hahn} J.~M.,
  {McGhee-French} C.~A.,  {Colwell} J.~E.,  {Marouf} E.~A.,   {Rappaport}
  N.~J.,  2016, \icarus, 279, 62

\bibitem[\protect\citeauthoryear{{French}, {McGhee-French}, {Nicholson}  \&
  {Hedman}}{{French} et~al.}{2019}]{French2019}
{French} R.~G.,  {McGhee-French} C.~A.,  {Nicholson} P.~D.,   {Hedman} M.~M.,
  2019, \icarus, 319, 599

\bibitem[\protect\citeauthoryear{{French}, {Bridges}, {Hedman}, {Nicholson},
  {Mankovich}  \& {McGhee-French}}{{French} et~al.}{2021}]{French2021}
{French} R.~G.,  {Bridges} B.,  {Hedman} M.~M.,  {Nicholson} P.~D.,
  {Mankovich} C.,   {McGhee-French} C.~A.,  2021, \mn@doi [\icarus]
  {10.1016/j.icarus.2021.114660}, \href
  {https://ui.adsabs.harvard.edu/abs/2021Icar..37014660F} {370, 114660}

\bibitem[\protect\citeauthoryear{{Fuller}}{{Fuller}}{2014}]{Fuller2014b}
{Fuller} J.,  2014, \icarus, 242, 283

\bibitem[\protect\citeauthoryear{{Fuller}, {Lai}  \& {Storch}}{{Fuller}
  et~al.}{2014}]{Fuller2014a}
{Fuller} J.,  {Lai} D.,   {Storch} N.~I.,  2014, \icarus, 231, 34

\bibitem[\protect\citeauthoryear{{Galanti} \& {Kaspi}}{{Galanti} \&
  {Kaspi}}{2021}]{Galanti2021}
{Galanti} E.,  {Kaspi} Y.,  2021, \mnras, 501, 2352

\bibitem[\protect\citeauthoryear{{Garc{\'\i}a-Melendo}, {P{\'e}rez-Hoyos},
  {S{\'a}nchez-Lavega}  \& {Hueso}}{{Garc{\'\i}a-Melendo}
  et~al.}{2011}]{Garcia-Melendo2011}
{Garc{\'\i}a-Melendo} E.,  {P{\'e}rez-Hoyos} S.,  {S{\'a}nchez-Lavega} A.,
  {Hueso} R.,  2011, \icarus, 215, 62

\bibitem[\protect\citeauthoryear{{Hachisu}}{{Hachisu}}{1986}]{Hachisu1986}
{Hachisu} I.,  1986, \mn@doi [\apjs] {10.1086/191121}, \href
  {https://ui.adsabs.harvard.edu/abs/1986ApJS...61..479H} {61, 479}

\bibitem[\protect\citeauthoryear{{Hedman} \& {Nicholson}}{{Hedman} \&
  {Nicholson}}{2013}]{Hedman2013}
{Hedman} M.~M.,  {Nicholson} P.~D.,  2013, \aj, 146, 12

\bibitem[\protect\citeauthoryear{{Hedman}, {Nicholson}  \& {French}}{{Hedman}
  et~al.}{2019}]{Hedman2019}
{Hedman} M.~M.,  {Nicholson} P.~D.,   {French} R.~G.,  2019, \aj, 157, 18

\bibitem[\protect\citeauthoryear{{Hubbard}}{{Hubbard}}{2013}]{Hubbard2013}
{Hubbard} W.~B.,  2013, \mn@doi [\apj] {10.1088/0004-637X/768/1/43}, \href
  {https://ui.adsabs.harvard.edu/abs/2013ApJ...768...43H} {768, 43}

\bibitem[\protect\citeauthoryear{{Jackson}, {MacGregor}  \&
  {Skumanich}}{{Jackson} et~al.}{2005}]{Jackson2005}
{Jackson} S.,  {MacGregor} K.~B.,   {Skumanich} A.,  2005, \mn@doi [\apjs]
  {10.1086/426587}, \href
  {https://ui.adsabs.harvard.edu/abs/2005ApJS..156..245J} {156, 245}

\bibitem[\protect\citeauthoryear{{Karino}}{{Karino}}{2003}]{Karino2003a}
{Karino} S.,  2003, \mn@doi [\mnras] {10.1046/j.1365-8711.2003.06649.x}, \href
  {https://ui.adsabs.harvard.edu/abs/2003MNRAS.343..175K} {343, 175}

\bibitem[\protect\citeauthoryear{{Karino} \& {Eriguchi}}{{Karino} \&
  {Eriguchi}}{2003}]{Karino2003b}
{Karino} S.,  {Eriguchi} Y.,  2003, \mn@doi [\apj] {10.1086/375768}, \href
  {https://ui.adsabs.harvard.edu/abs/2003ApJ...592.1119K} {592, 1119}

\bibitem[\protect\citeauthoryear{{Lai}}{{Lai}}{2001}]{Lai2001}
{Lai} D.,  2001, in {Centrella} J.~M.,  ed.,  American Institute of Physics
  Conference Series Vol. 575, Astrophysical Sources for Ground-Based
  Gravitational Wave Detectors. pp 246--257 (\mn@eprint {arXiv}
  {astro-ph/0101042}), \mn@doi{10.1063/1.1387316}

\bibitem[\protect\citeauthoryear{{Mankovich} \& {Fuller}}{{Mankovich} \&
  {Fuller}}{2021}]{Mankovich2021}
{Mankovich} C.~R.,  {Fuller} J.,  2021, \mn@doi [Nature Astronomy]
  {10.1038/s41550-021-01448-3}, \href
  {https://ui.adsabs.harvard.edu/abs/2021NatAs...5.1103M} {5, 1103}

\bibitem[\protect\citeauthoryear{{Mankovich}, {Marley}, {Fortney}  \&
  {Movshovitz}}{{Mankovich} et~al.}{2019}]{Mankovich2019}
{Mankovich} C.,  {Marley} M.~S.,  {Fortney} J.~J.,   {Movshovitz} N.,  2019,
  \mn@doi [\apj] {10.3847/1538-4357/aaf798}, \href
  {https://ui.adsabs.harvard.edu/abs/2019ApJ...871....1M} {871, 1}

\bibitem[\protect\citeauthoryear{{Marley}}{{Marley}}{1991}]{Marley1991}
{Marley} M.~S.,  1991, \mn@doi [\icarus] {10.1016/0019-1035(91)90239-P}, \href
  {https://ui.adsabs.harvard.edu/abs/1991Icar...94..420M} {94, 420}

\bibitem[\protect\citeauthoryear{{Marley} \& {Porco}}{{Marley} \&
  {Porco}}{1993}]{Marley1993}
{Marley} M.~S.,  {Porco} C.~C.,  1993, \mn@doi [\icarus]
  {10.1006/icar.1993.1189}, \href
  {https://ui.adsabs.harvard.edu/abs/1993Icar..106..508M} {106, 508}

\bibitem[\protect\citeauthoryear{{Nettelmann} et~al.,}{{Nettelmann}
  et~al.}{2021}]{Nettelmann2021}
{Nettelmann} N.,  et~al., 2021, \mn@doi [PSJ] {10.3847/PSJ/ac390a}, \href
  {https://ui.adsabs.harvard.edu/abs/2021PSJ.....2..241N} {2, 241}

\bibitem[\protect\citeauthoryear{{Ostriker} \& {Mark}}{{Ostriker} \&
  {Mark}}{1968}]{Ostriker1968}
{Ostriker} J.~P.,  {Mark} J.~W.~K.,  1968, \mn@doi [\apj] {10.1086/149506},
  \href {https://ui.adsabs.harvard.edu/abs/1968ApJ...151.1075O} {151, 1075}

\bibitem[\protect\citeauthoryear{{Passamonti} \& {Andersson}}{{Passamonti} \&
  {Andersson}}{2015}]{Passamonti2015}
{Passamonti} A.,  {Andersson} N.,  2015, \mn@doi [\mnras]
  {10.1093/mnras/stu2062}, \href
  {https://ui.adsabs.harvard.edu/abs/2015MNRAS.446..555P} {446, 555}

\bibitem[\protect\citeauthoryear{{Passamonti}, {Haskell}, {Andersson}, {Jones}
  \& {Hawke}}{{Passamonti} et~al.}{2009}]{Passamonti2009}
{Passamonti} A.,  {Haskell} B.,  {Andersson} N.,  {Jones} D.~I.,   {Hawke} I.,
  2009, \mn@doi [\mnras] {10.1111/j.1365-2966.2009.14408.x}, \href
  {https://ui.adsabs.harvard.edu/abs/2009MNRAS.394..730P} {394, 730}

\bibitem[\protect\citeauthoryear{{Reese}, {Ligni{\`e}res}  \&
  {Rieutord}}{{Reese} et~al.}{2006}]{Reese2006}
{Reese} D.,  {Ligni{\`e}res} F.,   {Rieutord} M.,  2006, \aap, 455, 621

\bibitem[\protect\citeauthoryear{{Reese}, {MacGregor}, {Jackson}, {Skumanich}
  \& {Metcalfe}}{{Reese} et~al.}{2009}]{Reese2009}
{Reese} D.~R.,  {MacGregor} K.~B.,  {Jackson} S.,  {Skumanich} A.,   {Metcalfe}
  T.~S.,  2009, \aap, 506, 189

\bibitem[\protect\citeauthoryear{{Reese}, {Prat}, {Barban}, {van 't
  Veer-Menneret}  \& {MacGregor}}{{Reese} et~al.}{2013}]{Reese2013}
{Reese} D.~R.,  {Prat} V.,  {Barban} C.,  {van 't Veer-Menneret} C.,
  {MacGregor} K.~B.,  2013, \aap, 550, A77

\bibitem[\protect\citeauthoryear{{Reese}, {Mirouh}, {Espinosa Lara}, {Rieutord}
   \& {Putigny}}{{Reese} et~al.}{2021}]{Reese2021}
{Reese} D.~R.,  {Mirouh} G.~M.,  {Espinosa Lara} F.,  {Rieutord} M.,
  {Putigny} B.,  2021, \mn@doi [\aap] {10.1051/0004-6361/201935538}, \href
  {https://ui.adsabs.harvard.edu/abs/2021A&A...645A..46R} {645, A46}

\bibitem[\protect\citeauthoryear{{Rieutord}, {Espinosa Lara}  \&
  {Putigny}}{{Rieutord} et~al.}{2016}]{Rieutord2016}
{Rieutord} M.,  {Espinosa Lara} F.,   {Putigny} B.,  2016, \mn@doi [Journal of
  Computational Physics] {10.1016/j.jcp.2016.05.011}, \href
  {https://ui.adsabs.harvard.edu/abs/2016JCoPh.318..277R} {318, 277}

\bibitem[\protect\citeauthoryear{{Schenk}, {Arras}, {Flanagan}, {Teukolsky}  \&
  {Wasserman}}{{Schenk} et~al.}{2002}]{Schenk2002}
{Schenk} A.~K.,  {Arras} P.,  {Flanagan} {\'E}.~{\'E}.,  {Teukolsky} S.~A.,
  {Wasserman} I.,  2002, \prd, 65, 024001

\bibitem[\protect\citeauthoryear{{Takata} \& {Saio}}{{Takata} \&
  {Saio}}{2013}]{Takata2013}
{Takata} M.,  {Saio} H.,  2013, \mn@doi [\pasj] {10.1093/pasj/65.3.68}, \href
  {https://ui.adsabs.harvard.edu/abs/2013PASJ...65...68T} {65, 68}

\bibitem[\protect\citeauthoryear{{Vorontsov}}{{Vorontsov}}{1981}]{Vorontsov1981}
{Vorontsov} S.~V.,  1981, \sovast, \href
  {https://ui.adsabs.harvard.edu/abs/1981SvA....25..724V} {25, 724}

\bibitem[\protect\citeauthoryear{{Wisdom} \& {Hubbard}}{{Wisdom} \&
  {Hubbard}}{2016}]{Wisdom2016}
{Wisdom} J.,  {Hubbard} W.~B.,  2016, \mn@doi [\icarus]
  {10.1016/j.icarus.2015.12.030}, \href
  {https://ui.adsabs.harvard.edu/abs/2016Icar..267..315W} {267, 315}

\bibitem[\protect\citeauthoryear{{Xu} \& {Lai}}{{Xu} \& {Lai}}{2017}]{Xu2017}
{Xu} W.,  {Lai} D.,  2017, \prd, 96, 083005

\makeatother
\end{thebibliography}




\appendix

\section{Model calculations}\label{app:model}
\begin{figure*}
    \centering
    \includegraphics[width=1.025\columnwidth]{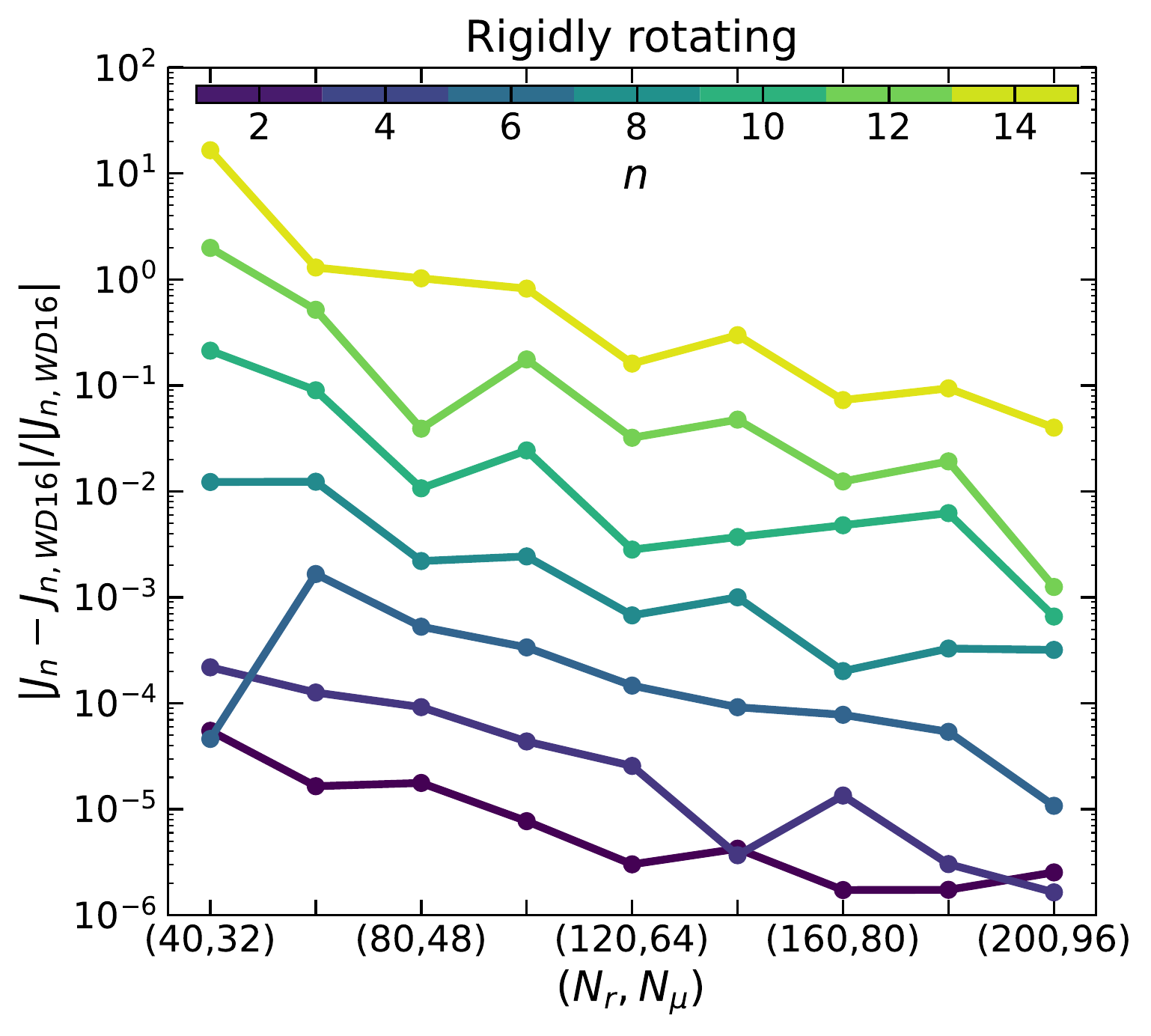}
    \includegraphics[width=.975\columnwidth]{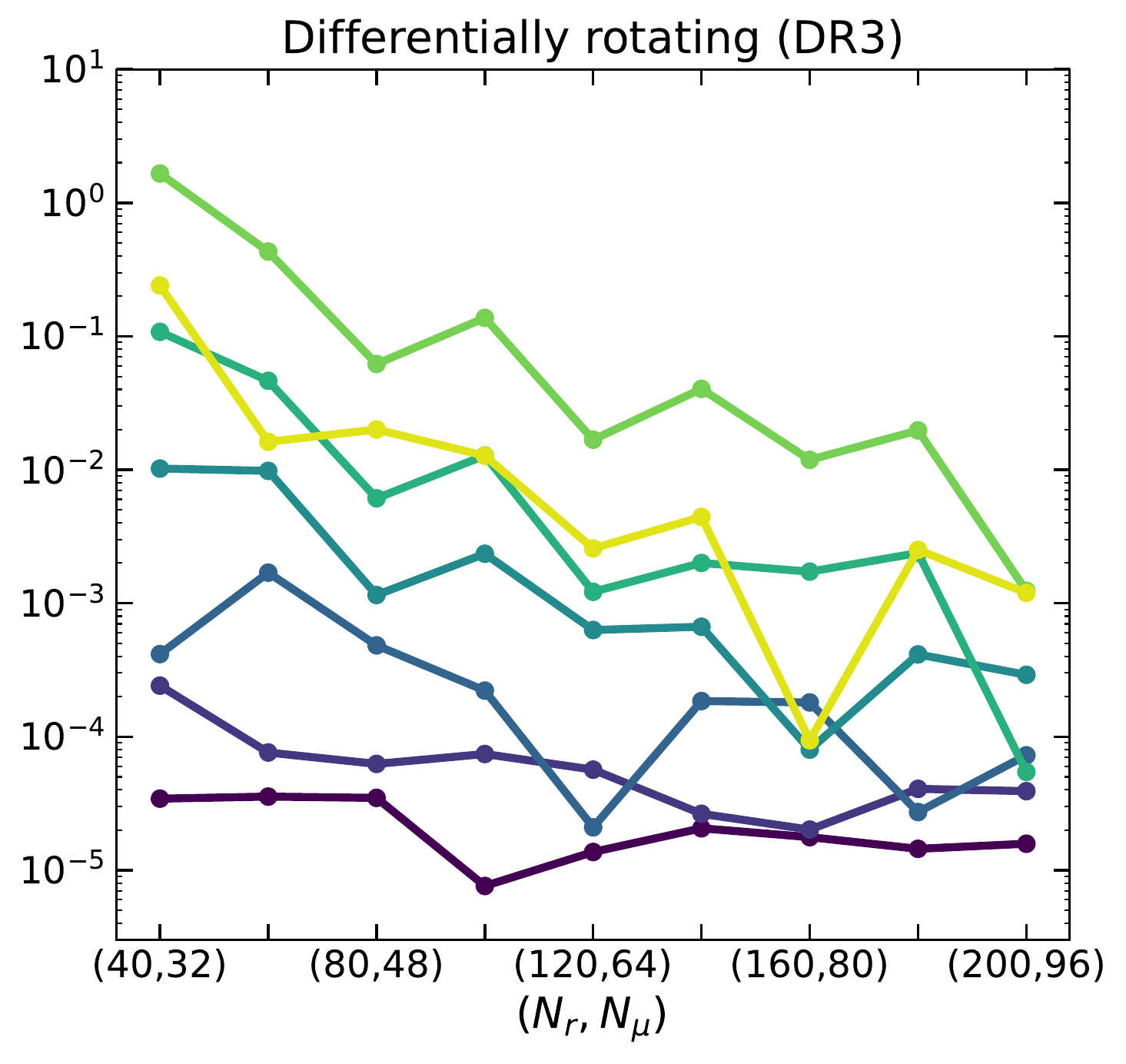}
    \caption{Left: relative deviations from values computed by \citet{Wisdom2016} for the gravitational moments $J_n$ of an $n=1$ polytrope with $(\Omega/\Omega_d)^2=0.089195487$, as a function of increasing radial and latitudinal resolutions $N_r$ and $N_\mu$. 
    Right: similar residuals, but computed for the ``DR3'' rotation profile of \citet{Wisdom2016}. For the differentially rotating models, we fix the central value of $\Omega$ relative to $\Omega_d$.}
    \label{fig:WD16cpr}.
\end{figure*}
\begin{figure}
    \centering
    \includegraphics[width=\columnwidth]{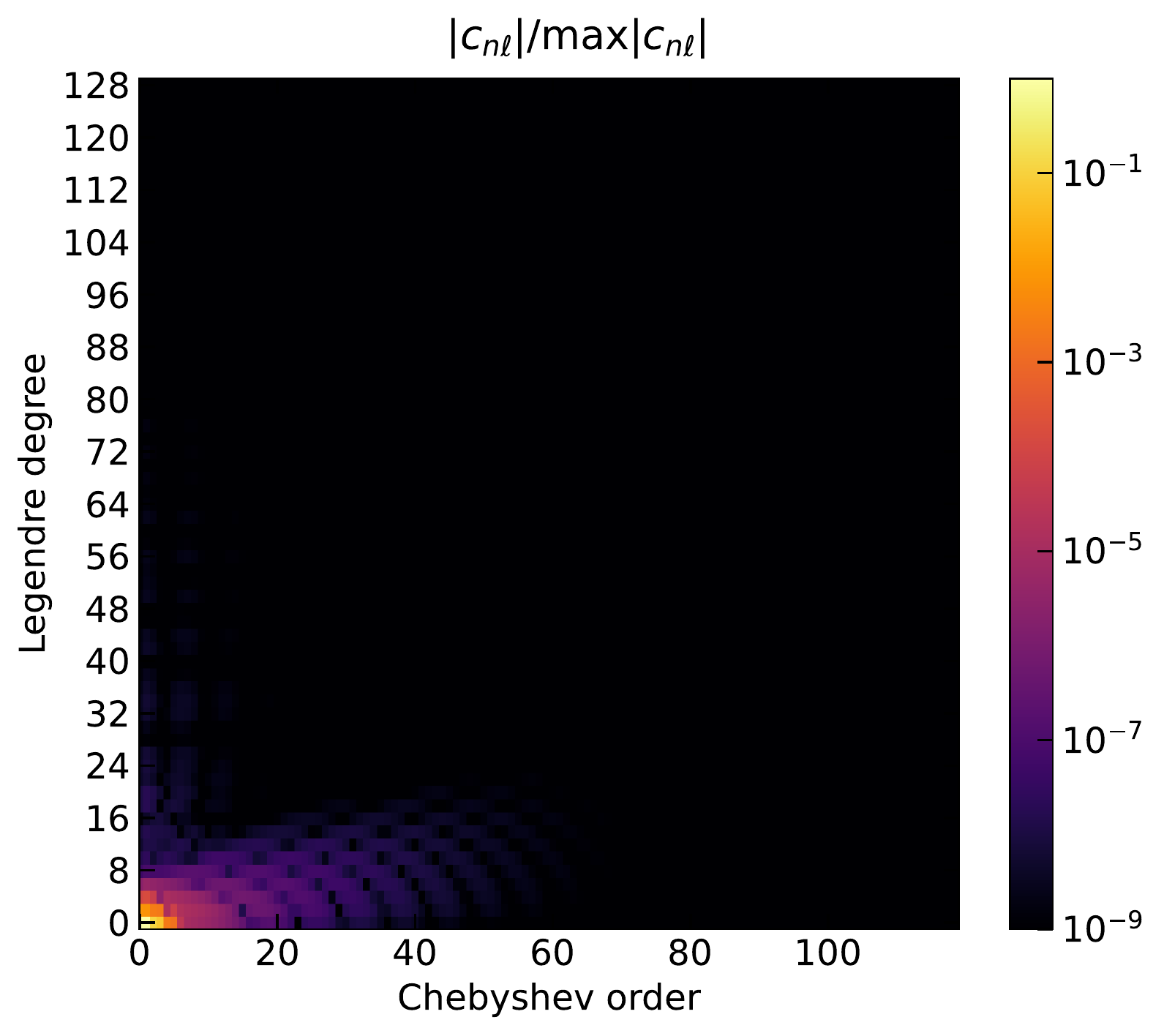}
    \caption{Envelope of spectral coefficients in the expansion of Equation \eqref{eq:ThExp} for the differentially rotating model characterized by \autoref{fig:WD16cpr} (right), computed with $N_r=120,N_\mu=64.$}
    \label{fig:drot_cf}
\end{figure}

This appendix describes results from our polytropic model calculations that we have compiled for comparison with previous work. In the planetary sciences community, rotating equilibria are often characterized in terms of gravitational moments $J_n$ that appear in expansions of the (external) potential with the form
\begin{equation}
    \Phi_0(r,\mu)=\frac{GM}{R_\text{eq}}
    \sum_{\text{n}=0}^\infty
    \left(
        \frac{R_\text{eq}}{r}
    \right)^{\text{n}+1}J_\text{n}P_\text{n}(\mu),
\end{equation}
where the $P_\text{n}$ are Legendre polynomials. These $J_\text{n}$ coefficients can be computed with numerical quadratures via
\begin{equation}\label{eq:Jn1}
    J_\text{n}=\frac{-1}{MR_\text{eq}^\text{n}}
    \int_V\rho_0 r^\text{n}P_\text{n}\text{d}V,
\end{equation}
or from the expansion $\sum_\ell \Phi^\ell(r)Y_\ell^{m=0}$ in zonal spherical harmonics that is a by-product of our model calculations:
\begin{equation}\label{eq:Jn2}
    J_\text{n}
    =\left(\frac{R_\text{eq}}{GM}\right)
    \left[\frac{(2\text{n}+1)}{4\pi}\right]^{1/2}
    \left(\frac{r}{R_\text{eq}}\right)^{\text{n}+1}
    \Phi^\text{n}(r).
\end{equation}
We have verified that \autoref{eq:Jn1} and \autoref{eq:Jn2} give nearly identical results (since solutions satisfy $\Phi^\text{n}\propto r^{-\text{n}-1}$ in vacuum).

\autoref{fig:WD16cpr} (left) shows relative differences between $J_2-J_{14}$ coefficients that we have computed (with the approach described in \autoref{sec:eqm}) for a rigidly rotating $n=1$ polytrope rotating with  $(\Omega/\Omega_d)^2=0.089195487$, as compared with values obtained semi-analytically by \citet{Wisdom2016}. \autoref{fig:WD16cpr} (right) plots the same differences in $J_n$ coefficients, but computed for the ``DR3'' profile of differential rotation described by the same authors. In both cases, deviations for our working resolution of $N_r=120,N_\mu=64$ compare favorably to those obtained by \citet{Nettelmann2021} with seventh order theory of figures and $10^4-10^5$ grid points (cf. their fig. 1). Relative errors in higher-$\text{n}$ coefficients are large because the corresponding coefficients are very small.

Regardless of the convergence of the $J_\text{n}$-coefficients, precise values of which are more important for comparing with observations than computing modes, the spectral expansions determining the equilibrium structures in our models appear well-resolved. \autoref{fig:drot_cf} shows the ``envelope'' of spectral coefficients in the expansion introduced in \autoref{eq:ThExp}, computed with $N_r=120,N_\mu=64$ for the same differentially rotating, $n=1$ polytrope. Beyond Chebyshev order $\simeq60,$ and harmonic degree $\simeq20,$ coefficient values fall to less than a billionth of the maximum value.

\begin{figure*}
    \centering
    \includegraphics[width=\textwidth]{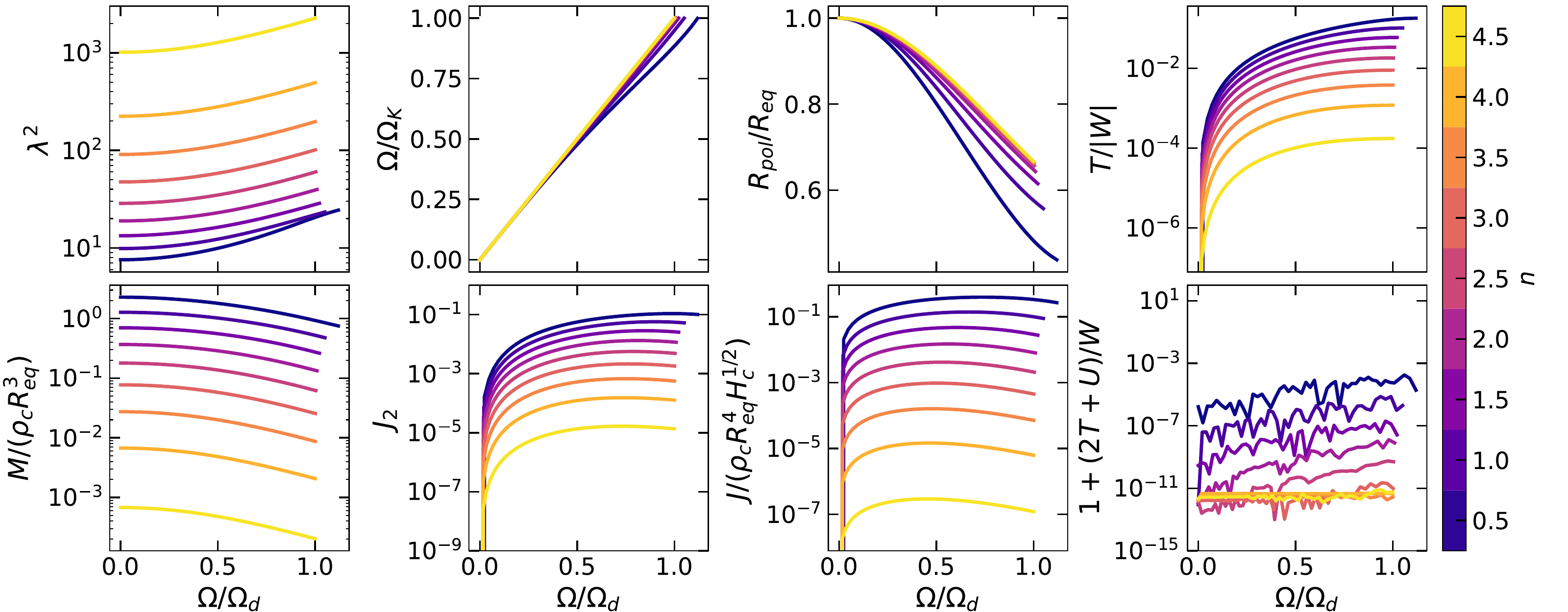}
    \caption{Plots showing the eigenvalues $\lambda^2$ (top left), rotation rates relative to the equatorial surface Keplerian rate (top middle left; $\Omega_K^2=r^{-1}\partial_r\Phi$), ratio of polar to equatorial radii (top middle right), ratios of total kinetic to gravitational potential energy (top right), total masses normalized by central densities (bottom left), quadrupolar gravitational moments (bottom middle left), total angular momenta (bottom middle right), and virial errors (bottom right) computed for rigidly rotating models with rotation rates extending from zero to the mass-shedding limit (where $\Omega=\Omega_K$ at the equator). The colormap extends from polytropic indices $n=0.5$ (purple) to $n=4.5$ (yellow). For lower polytropic indices, note that the mass shedding limit extends beyond the dynamical frequency $\Omega_d=(GM/R_\text{eq}^3)^{1/2}$.}
    \label{fig:model_grid}
\end{figure*}

\begin{figure*}
    \centering
    \includegraphics[width=\textwidth]{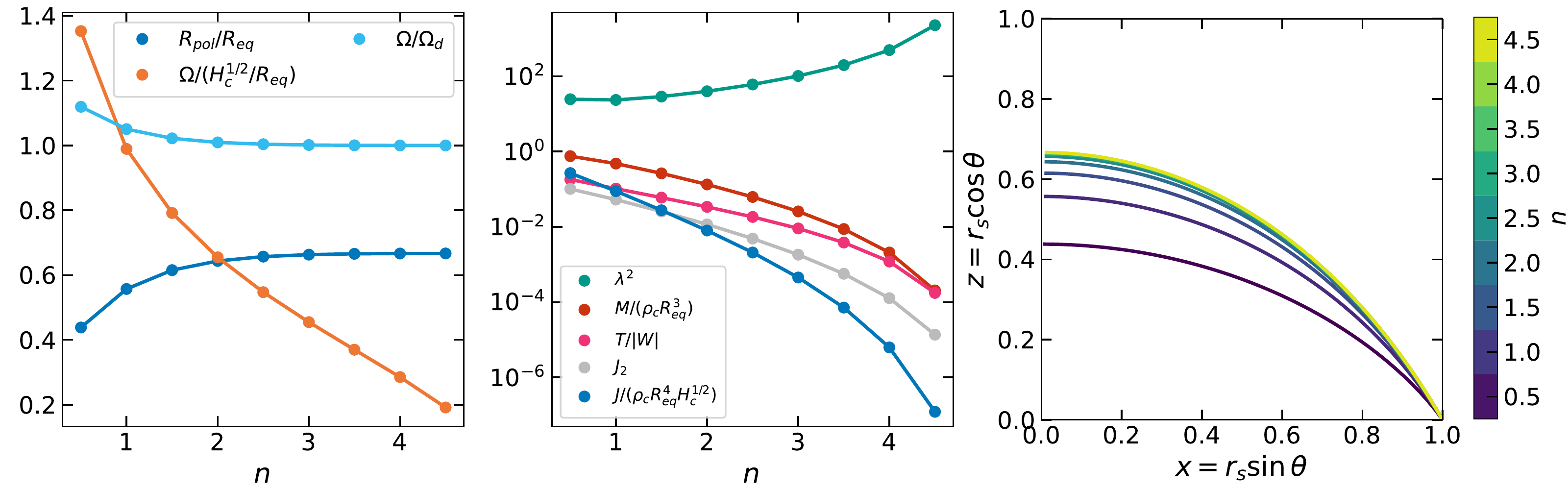}
    \caption{Left, middle: model properties for rigidly rotating polytropes at the critical, mass-shedding limit. Right: plots showing the oblate surface of the critically rotating models (polytropic indices again range from $n=0.5$ to $n=4.5$).}
    \label{fig:crit_rot}
\end{figure*}

\autoref{fig:model_grid} displays relevant quantities for rigidly rotating models with polytropic indices from $n=0.5$ (blue) to $n=4.5$ (yellow), and rotation rates extending to just below the critical ``mass-shedding'' limit determined by exact balance between the centrifugal and gravitational accelerations at the equator, i.e., 
$\Omega=\Omega_K,$ 
where $\Omega_K^2=r^{-1}\partial_r\Phi_0|_{r=R_\text{eq},\mu=0}$. As shown in \autoref{fig:model_grid} (top middle left), for large polytropic indices $\Omega_K\simeq\Omega_d$. For lower $n\lesssim1,$ however, the non-spherical gravitational field associated with stronger flattening of the equilibrium structure (see \autoref{fig:model_grid}, top middle right) can lead to critical rotation rates marginally larger than the dynamical frequency $\Omega_d$. \autoref{fig:crit_rot} illustrates the dependence of model attributes on index $n$ for polytropes with $\Omega/\Omega_K\simeq1$.

Calculating the total mass, angular momentum, rotational kinetic energy, internal energy, and gravitational energy as
\begin{align}
    M&=\int_V\rho_0\text{d}V, \\
    J&=\int_V\rho_0R^2\Omega\text{d}V,\\
    T&=\frac{1}{2}\int_V\rho_0R^2\Omega^2\text{d}V,\\
    U&=3\int_VP_0\text{d}V, \\
    W&=\frac{1}{2}\int_V\rho_0\Phi_0\text{d}V,
\end{align}
we find values in agreement with previous results, for both rigidly and differentially rotating models \citep{Hachisu1986,Eriguchi1985,Passamonti2009,Passamonti2015}. \autoref{fig:model_grid} (bottom right) also plots virial errors $\epsilon_V=1 + (2T+U)/W,$ which test convergence by assessing the degree to which rotating equilibria satisfy the virial theorem \citep[e.g.,][]{Rieutord2016}. For the $n\geq1$ polytropes considered in this paper, we achieve $\epsilon_V\lesssim10^{-6}$.

\section{Mode calculations}\label{app:modes}
In this appendix we outline the numerical methods used to compute modes, which largely follow those of \citet{Reese2006,Reese2009}. Expressed on the dual (contravariant) basis of a curvilinear coordinate system with Christoffel symbols satisfying $\Gamma^i_{jk}=\Gamma^i_{kj},$ Equations \eqref{eq:l1}-\eqref{eq:l4} can be written as
\begin{align}
    \text{i}\sigma g_{ij}v^j
    &=g_{ij}(
        u_0^k\partial_kv^j
        +2\Gamma^j_{kl}v^ku_0^l
        +v^k\partial_k u_0^j
    )
    \\\notag&
    +\rho_0^{-1}[
        \partial_i(\rho_0h) 
        -G_i\delta\rho
    ]
    +\partial_i\delta\Phi,
\\
    d_t\delta\rho
    &=-J^{-1}\partial_j(J\rho_0v^j),
\\
    d_t(\rho_0 h)-c_A^2d_t\delta\rho
    &=-\rho_0v^jA_j,
\\
    0&=4\pi G\delta\rho
    -J^{-1}\partial_j(Jg^{jk}\partial_k\delta\Phi).
\end{align}
Here upper (lower) indices denote the contravariant (covariant) components associated with covariant (contravariant) basis vectors, $\partial_i$ is the partial derivative with respect to the $i$'th coordinate, $g_{ij}$ is the metric tensor, $g^{ij}$ is the inverse metric tensor, $J=(\text{det}\  g_{ij})^{1/2}$ is the Jacobian, and paired upper and lower indices denote summation. We have written $d_t=\partial_t + u_0^j\partial_j$, as well as $A_i=\rho_0^{-1}\partial_i P_0-c_A^2\partial_i\ln\rho_0$, and introduced the linear enthalpy perturbation $h=\delta P/\rho_0$. Depending on the equilibrium state, it can be advantageous to trade $\delta\rho$ for the variable $\delta\rho/\rho_0$; we use $\delta\rho$ for our calculations employing $n=1$ polytropes, and $\delta\rho/\rho_0$ for $n=1.6.$

Like \citet{Reese2006,Reese2009,Reese2013,Reese2021} and \citet{Dewberry2021}, we use a non-orthogonal coordinate system $(\zeta,\theta,\phi)$ with quasi-radial coordinate $\zeta$ defined (in units with $R_\text{eq}=1$) by the mapping \citep[originally proposed by ][]{Bonazzola1998}
\begin{equation}\label{eq:rzt1}
    r(\zeta,\theta)
    =(1-\epsilon)\zeta+\frac{1}{2}(5\zeta^3-3\zeta^5)(r_s-1+\epsilon)
\end{equation}
for $\zeta\in[0,1]$, and
\begin{equation}\label{eq:rzt2}
    r(\zeta,\theta)
    =2\epsilon+(1-\epsilon)\zeta+(2\zeta^3-9\zeta^2+12\zeta-4)(r_s-1-\epsilon)
\end{equation}
for $\zeta\in[1,2]$. Here $r_s(\theta)$ is the surface of the oblate model, and $\epsilon=1-R_\text{pol}/R_\text{eq}$ characterizes centrifugal flattening. Equations \eqref{eq:rzt1} and \eqref{eq:rzt2} imply that $\zeta$ equals one on the stellar/planetary surface, and relaxes to spherical radius at $\zeta=0$ and $\zeta=2$. The outer vaccum $\zeta\in[1,2]$ is included for the purpose of applying boundary conditions on the gravitational potential.

The natural covariant basis vectors ${\bf E}_i=\partial_i{\bf r}$ associated with $(\zeta,\theta,\phi)$ coordinates are related to the unit spherical basis by \cite[e.g.,][]{Rieutord2016}
\begin{align}
    {\bf E}_\zeta&=\partial_\zeta r\hat{\bf r},\\
    {\bf E}_\theta&=\partial_\theta r\hat{\bf r}
    +r\hat{\boldsymbol{\theta}},\\
    {\bf E}_\phi&=r\sin\theta\hat{\boldsymbol{\phi}}.
\end{align}
The equilibrium velocity fields ${\bf u}_0=r\sin\theta\Omega\hat{\boldsymbol{\phi}}$ considered in this paper therefore take the simple form $u_0^i{\bf E}_i=\Omega{\bf E}_\phi$, and hence $d_t=-\text{i}\sigma + \Omega\partial_\phi.$ 

Assuming that perturbations adopt a harmonic dependence in azimuth as well as time (e.g., writing $\delta\rho\propto\exp[i(m\phi-\sigma t)]$), inserting nonzero geometric factors \citep[provided in, e.g.,][]{Rieutord2016}, and writing $s=\sin\theta,$ $\mu=\cos\theta$, $r_i=\partial_i r$,  the linearized equations are (in units with $G=M=R_\text{eq}=1$)
\begin{align}
    \text{i}\omega&\left(
        r_\zeta^2v^\zeta 
        +r_\zeta r_\theta v^\theta
    \right)
    =-2\Omega r_\zeta r s^2 v^\phi
    \\\notag&\hspace{4em}
    +\rho_0^{-1}\left[\partial_\zeta(\rho_0 h)-G_\zeta\delta\rho\right]
    +\partial_\zeta\delta\Phi,
\\
    \text{i}\omega&\left[
        r_\zeta r_\theta v^\zeta 
        +\left(r^2 + r_\theta^2\right)v^\theta
    \right]
    =-2\Omega r s(r\mu + r_\theta s)v^\phi
    \\\notag&\hspace{4em}
    +\rho_0^{-1}\left[
        \partial_\theta(\rho_0 h)
        -G_\theta\delta\rho 
    \right]
    +\partial_\theta\delta\Phi,
\\
    \text{i}\omega& v^\phi
    =\left(
        \partial_\zeta\Omega 
        +2\Omega \frac{r_\zeta}{ r}
    \right)v^\zeta
    \\\notag&\hspace{4em}
    +\left[ 
        \partial_\theta\Omega 
        + 2\Omega\left(
            \frac{r_\theta s + r\mu}{rs}
        \right)
    \right]v^\theta
    +\frac{\text{i}m}{r^2s^2}(h + \delta\Phi),
\\
    \text{i}\omega& \delta\rho
    =\frac{1}{r_\zeta r^2}\partial_\zeta \left(r_\zeta r^2\rho_0 v^\zeta\right)
    \\\notag&\hspace{4em}
    +\frac{1}{r_\zeta r^2s}\partial_\theta \left(r_\zeta r^2s\rho_0 v^\theta\right)
    +\text{i}m\rho_0v^\phi,
\\
    \text{i}\omega& (
        \rho_0h
        -c_A^2\delta\rho
    )
    =\rho_0(A_\zeta v^\zeta + A_\theta v^\theta),
\\
    0&=4\pi r^2\delta \rho-
    \Bigg\{
        \left(\dfrac{r^2+r_{\theta}^2}{r_\zeta^2}\right)
        \partial^2_{\zeta\zeta}
        \\\notag&\hspace{4em}
        +\left(
            2\frac{r}{r_\zeta}
            +2\frac{r_\theta r_{\zeta\theta}}{r_\zeta^2}
            -(r^2+r_\theta^2)\frac{r_{\zeta\zeta}}{r_\zeta^3}
            -\frac{r_{\theta\theta}}{r_\zeta}
            -\frac{\mu r_\theta}{s r_\zeta}
        \right)\partial_\zeta
        \\\notag&\hspace{4em}
        -\dfrac{2r_{\theta}}{r_\zeta} \partial^2_{\theta\zeta}
        +\partial^2_{\theta\theta}
        +\frac{\mu}{s}\partial_\theta
        -\dfrac{m^2}{s^2}
    \Bigg\}\delta\Phi,
\end{align}
where $\omega=\sigma-\text{i}m\Omega$. For rigidly rotating bodies $\omega$ is the mode frequency in the corotating frame; for differentially rotating bodies, such a frame is not well-defined. 

Introducing the expansions 
\begin{align}
    v^\zeta(\zeta,\theta,\phi)&=\frac{\zeta^2}{r^2r_\zeta}\sum_{\ell=|m|}^\infty a^\ell(\zeta)Y_\ell^m,
\\
    v^\theta(\zeta,\theta,\phi)
    &=\frac{\zeta}{r^2r_\zeta}\sum_{\ell=|m|}^\infty 
    \left[
        b^\ell(\zeta)\partial_\theta Y_\ell^m
        +c^\ell(\zeta)D_\phi Y_\ell^m
    \right],
\\
    v^\phi(\zeta,\theta,\phi)
    &=\frac{\zeta}{r^2r_\zeta s}\sum_{\ell=|m|}^\infty 
    \left[ 
        b^\ell(\zeta)D_\phi Y_\ell^m
        - c^\ell(\zeta)\partial_\theta Y_\ell^m
    \right],
\\
    \delta\rho(\zeta,\theta,\phi)
    &=\sum_{\ell=|m|}^\infty \rho^\ell(\zeta)Y_\ell^m,
\\
    h(\zeta,\theta,\phi)
    &=\sum_{\ell=|m|}^\infty h^\ell(\zeta)Y_\ell^m,
\\
    \delta\Phi(\zeta,\theta,\phi)
    &=\sum_{\ell=|m|}^\infty \Phi^\ell(\zeta)Y_\ell^m,
\end{align}
where $D_\phi=s^{-1}\partial_\phi$, we follow \citet{Reese2006} in projecting onto spherical harmonics. This produces an infinite set of $\zeta-$dependent equations that are coupled by both the Coriolis force, and geometric factors associated with the non-orthogonal coordinate system. We solve these coupled ordinary differential equations in $\zeta$ using pseudospectral collocation with Chebyshev Cardinal functions \citep{Boyd2001}. As described in \citet{Dewberry2021}, we use boundary bordering to enforce (i) regularity at $r=\zeta=0$, (ii) a vanishing Lagrangian pressure perturbation and a continuous gravitational potential at $\zeta=1$ (the planetary surface), and (iii) the matching of the gravitational potential to solutions that vanish at infinity at $\zeta=2.$ 

In the $\zeta$-direction we compare calculations on Gauss-Lobatto grids with $N_\zeta=90$ and $120$ collocation points for validation (typically, differences in frequency between the two are $\lesssim 1$ppm). In the $\mu$-direction, we compute projection integrals on a Gauss-Legendre grid with $N_\mu=64$ points in the half-plane, and truncate perturbations' spherical harmonic expansions at a maximum degree $\ell_\text{max}$ such that the maximum values of the spectral coefficients $a^{\ell_\text{max}},b^{\ell_\text{max}},c^{\ell_\text{max}},\rho^{\ell_\text{max}},h^{\ell_\text{max}},\Phi^{\ell_\text{max}}$ are all at least $100$ times smaller than the maximum values of any of the coefficients in the corresponding expansions. For most of the modes considered, $\ell_\text{max}\leq m+40$ is more than sufficient.

\section{Validation of mode calculations}\label{app:barmode}

\begin{figure*}
    \centering
    \includegraphics[width=\textwidth]{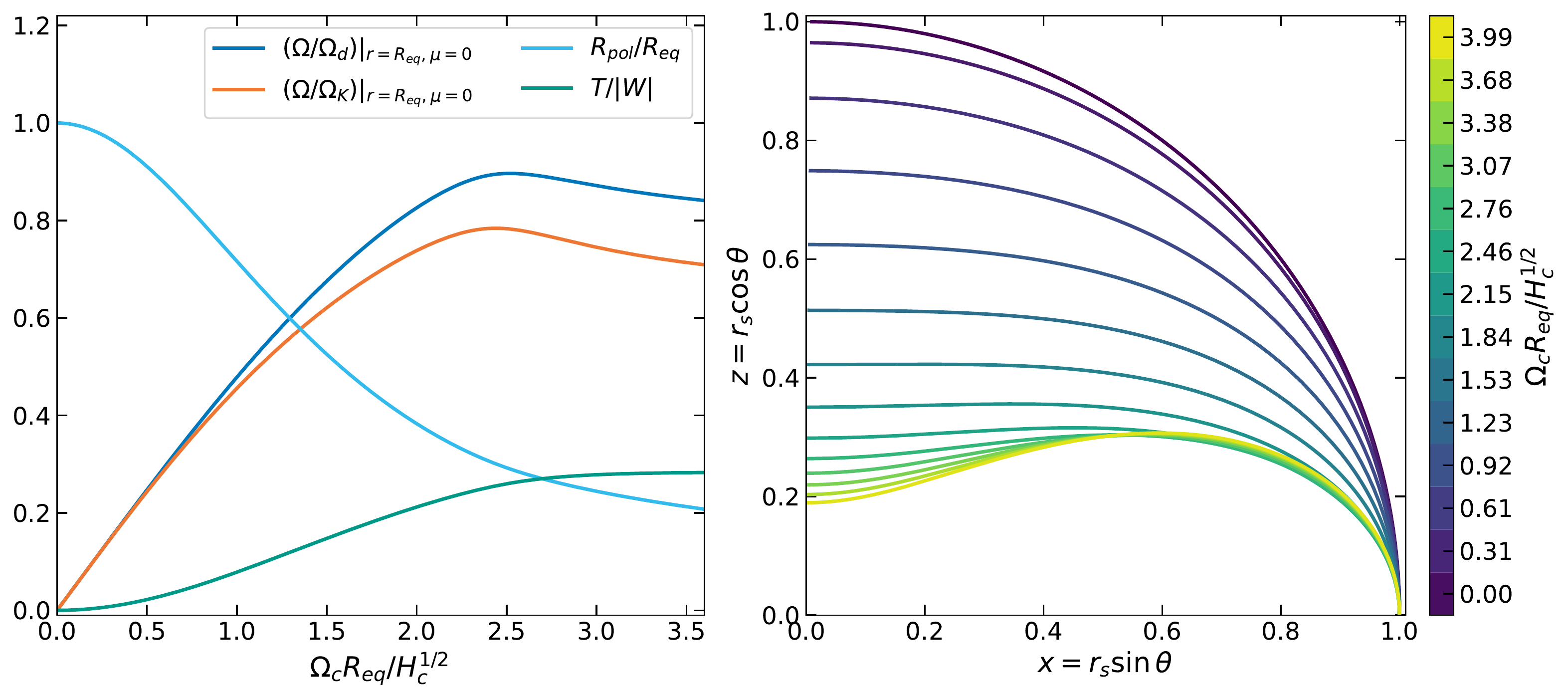}
    \caption{Model parameters for a sequence of differentially rotating, $n=1$ polytropes with a ``constant-j'' profile for angular velocity. The left-hand panel plots the surface equatorial value of the rotation rate relative to both the dynamical frequency $\Omega_d=(GM/R_{eq}^3)^{1/2}$ (dark blue) and the Keplerian frequency $\Omega_K=(r^{-1}\partial_r\Phi)^{1/2}$ (orange), the ratio of polar to equatorial radii (light blue), and the ratio of kinetic to gravitational potential energy (teal) as a function of the parameter $\Omega_c$ (in units used for model computations). The right-hand panel shows profiles of the surface radius $r_s(\mu)$, with the color-scale indicating different values of $\Omega_c$.}
    \label{fig:cstj_models}
\end{figure*}

There is, to our knowledge, a limited selection of published benchmark oscillation mode computations for rapidly and \emph{differentially} rotating models of planets and stars. However, we have closely reproduced the results of previous calculations \citep{Karino2003a,Karino2003b,Passamonti2015} of dynamically unstable f-modes in polytropes rotating with ``constant-j'' rotation laws \citep{Hachisu1986,Eriguchi1985} with the form
\begin{equation}
    \Omega(R)=\frac{\Omega_c A^2}{R^2+A^2}.
\end{equation}
Here $A$ is a parameter (given in units of $R_\text{eq}$) that controls the degree of differential rotation, and $\Omega_c$ the overall rotation rate. \autoref{fig:cstj_models} (left) plots relevant quantities as a function of $\Omega_c$ for $n=1$ polytropes with a constant-j rotation laws and $A=1$, while \autoref{fig:cstj_models} (right) shows the corresponding changes in surface radius; because the differential rotation is strongest near the axis of rotation, the polytrope is most significantly flattened near $R=0.$  

\begin{figure*}
    \centering
    \includegraphics[width=\textwidth]{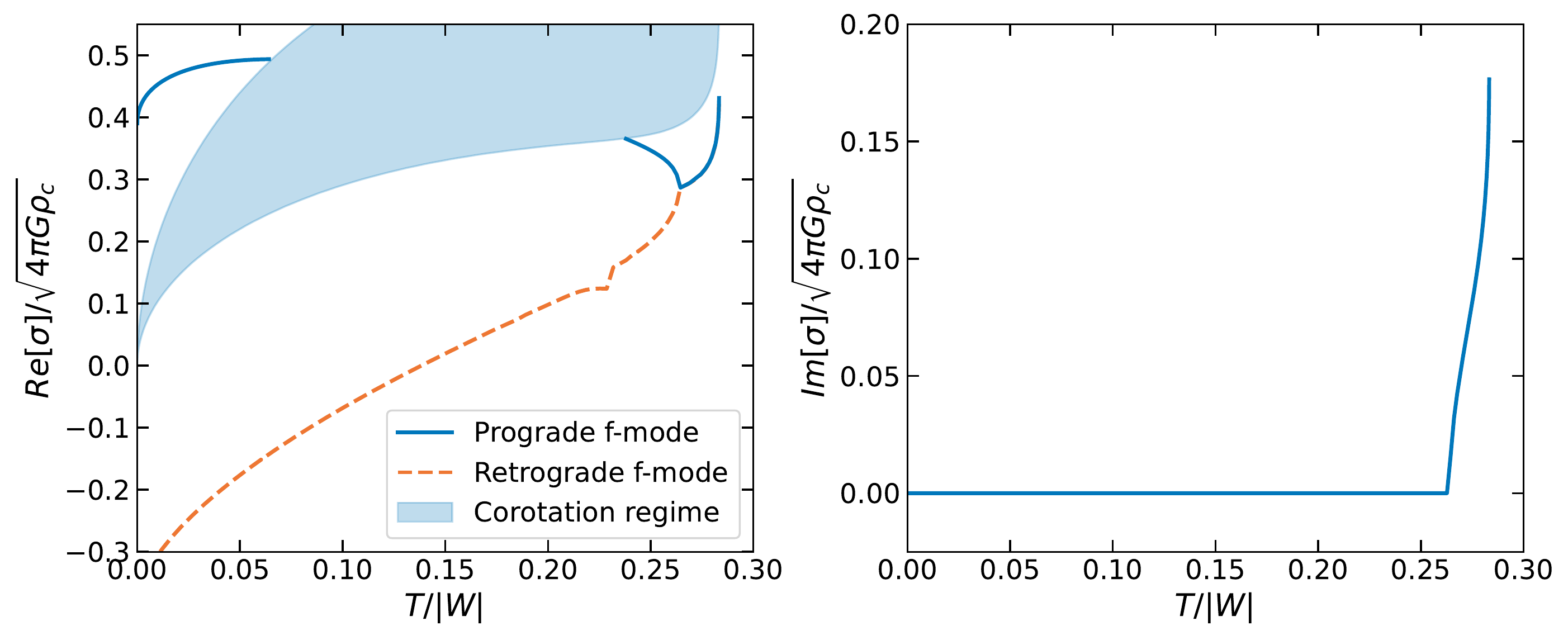}
    \caption{Sectoral $m=2$ f-mode frequencies (left) and growth rates (right) as a function of $T/|W|$ for the sequence of models described by Fig. \ref{fig:cstj_models}. For $T/|W|\gtrsim0.264,$ dynamical instability sets in as the prograde and retrograde f-modes become complex conjugate pairs (modes with identical frequencies, and growth rates with the same amplitude and opposite sign). The curves appear identical to those of \citet{Karino2003b}, save for a jump in the frequency of the retrograde f-mode at $T/|W|\simeq0.23$ that occurs because of an avoided crossing with a long-wavelength inertial mode.}
    \label{fig:cstj_modecpr}
\end{figure*}

Such profiles of differential rotation permit ratios of total kinetic to gravitational energy exceeding $\simeq0.27$ (unlike purely rigid rotation, for all but very low polytropic indices; see \autoref{fig:model_grid}, top right). This leads to dynamical ``bar-type'' instabilities involving $m=2$ f-modes with exponentially growing amplitudes \citep[e.g.,][]{Lai2001}. \autoref{fig:cstj_modecpr} illustrates this transition to instability, plotting the real (left) and imaginary (right) parts of the frequencies of prograde and retrograde, sectoral, $m=2$ f-modes computed for the models shown in \autoref{fig:cstj_models}. At a critical value of $T/|W|\simeq0.264,$ the real parts of the frequencies coincide, at which point the oscillations branch into a conjugate pair of exponentially growing and decaying modes.

\autoref{fig:cstj_modecpr} closely reproduces Figs. 1 and 2 in \citet{Karino2003b}, except that we do not attempt to resolve the prograde f-mode as it passes through the (blue-shaded) regime in which it possesses a corotation resonance inside the star. The jump in frequency of the retrograde f-mode near $T/|W|\simeq0.23$ occurs due to an avoided crossing with a long-wavelength inertial mode \citep[see, e.g.,][]{Dewberry2022}.


\bsp	
\label{lastpage}
\end{document}